\input{aipcheck}

\documentclass[
  ,draft            % use draft while you are working on the paper
  ,numberedheadings % uncomment this option for numbered sections
  ]
  {aipproc}

\layoutstyle{6x9}

\usepackage{amstext}
\usepackage{amssymb}
\usepackage{amsxtra}

\newcommand{\mub}[0]{\mu_{\mathrm{B}}}
\newcommand{\cdd}[0]{C_{\rm dd}}
\newcommand{\edd}[0]{\varepsilon_{\rm dd}}
\newcommand{\udd}[0]{U_{\rm dd}}
\newcommand{\fdd}[0]{\Phi_{\rm dd}}
\newcommand{\ket}[1]{|#1\rangle}

\usepackage{color}
\definecolor{gris}{rgb}{.9,.9,.9}
\newcounter{pbnb}
\begin{document}

\title{Dipolar interaction in ultra-cold atomic gases}



\author{C. Menotti}{
  address={ICFO -- Institut de Ci\`encies
Fot\`oniques, E-08860 Castelldefels, Barcelona, Spain \\
and CNR-INFM-BEC and Dipartimento di Fisica, 
Universit\`a di Trento, I-38050 Povo, Italy}
}

\author{M. Lewenstein}{
  address={ICFO -- Institut de Ci\`encies
Fot\`oniques, E-08860 Castelldefels, Barcelona, Spain \\
 and ICREA -- Instituci\'o Catalana de Recerca i Estudis 
Avan\c cats,  E-08010 Barcelona, Spain}
}

\author{T. Lahaye}{
  address={5. Physikalisches Institut, Universit\"at Stuttgart,
D-70550 Stuttgart, Germany} 
}

\author{T. Pfau}{
  address={5. Physikalisches Institut, Universit\"at Stuttgart,
D-70550 Stuttgart, Germany} 
}

\begin{abstract}
Ultra-cold atomic systems provide a new setting where to investigate
the role of long-range interactions. In this paper we will review the
basics features of those physical systems, in particular focusing on
the case of Chromium atoms.
On the experimental side, we report on the observation of dipolar
effects in the expansion dynamics of a Chromium Bose-Einstein
condensate.  By using a Feshbach resonance, the scattering length
characterising the contact interaction can be strongly reduced, thus
increasing the relative effect of the dipole-dipole interaction. Such
experiments make Chromium atoms the strongest candidates at present
for the achievement of the strong dipolar regime.
On the theoretical side, we investigate the behaviour of ultra-cold
dipolar systems in the presence of a periodic potential. We discuss
how to realise this situation experimentally and we characterise the
system in terms of its quantum phases and metastable states,
discussing in detail the differences with respect to the case of
zero-range interactions.

\end{abstract}

\maketitle

%%%%%%%%%%%%%%%%%%%%%%%%%%%%%%%%%%%%%%%%%%%%
%% MAINMATTER
%%%%%%%%%%%%%%%%%%%%%%%%%%%%%%%%%%%%%%%%%%%%

\section{Introduction: ultra-cold atomic systems}

The cooling and trapping techniques for neutral atoms developed in the
last 20 years have allowed the achievement of Bose-Einstein
condensation \cite{bec123}, quantum degenerate Fermi gases \cite{jin},
and the creation of several kinds of correlated systems. In all those
cases, a crucial role is played by the statistics of the atoms, the
interatomic interactions, the geometry of the system and their
interplay.

The first experiments were carried out with ultra-cold bosons in the
weakly interacting regime, where the Gross-Pitaevskii equation
provides a good description of the system.  Due to the very low
temperature in those systems, interactions are usually s-wave and
hence well described by a zero-range contact potential.  Within the
Gross-Pitaevskii theory, a huge variety of phenomena has been
described, like collective oscillations \cite{coll_osc}, interference
\cite{interference} and coherence effects \cite{coherence}, 
non linear atom optics (4-wave mixing \cite{nonlinear} and solitons
\cite{solitons}), superfluidity (sound propagation \cite{sound}, 
scissor modes \cite{scissor}, and quantised vortices
\cite{vortices}). In polarised (single species) fermionic
systems s-wave interactions are forbidden due to the Pauli principle,
so that interactions are normally introduced by considering two
different internal sublevels, mixtures of different species or p-wave
interactions.

By increasing the strength of the interactions (e.g. by enhancing the
scattering length via Feshbach resonances \cite{feshbach}) or by
changing the dimensionality of the system, correlations can be
introduced. Striking examples in this direction were the observation
of the superfluid to Mott insulator transition in optical lattices
\cite{greiner} and the Berezinskii-Kosterlitz-Thouless transition 
\cite{BKT},  the realisation of a Tonks-Girardeau gas
\cite{tonks}, the study of the BCS-BEC crossover in fermionic
systems \cite{bcsbec} and the creation of ultra-cold molecules
\cite{molecules}.

A further possibility is to consider atoms interacting via long-range
dipolar interactions ~\cite{baranov}. This will be the topic of this
paper, which summarises the talks given at the conference {\it
Dynamics and thermodynamics of systems with long range interaction:
theory and experiments} (Assisi, Italy -- July 2007) by Chiara Menotti
from ICFO, Barcelona and Tobias Koch from the group of Tilman Pfau at
the University of Stuttgart. We will explain how long-range
interactions arise in ultra-cold atomic systems, and in particular
present the observation of weak and strong dipolar interactions in a
degenerate sample of Chromium atoms. Moreover, we will discuss the
effect of dipolar interactions on atoms trapped in an optical lattice
and point out the striking differences with respect to the case of
zero-range interactions. The most recent experimental and theoretical
results presented here are contained respectively in our papers
\cite{nature,menotti}.

For the interested reader, extensive reviews on the physics of
ultra-cold bosonic and fermionic ultra-cold gases can be found in
\cite{dalfovo, bec, pethick, bloch, giorgini}.

\section{Interatomic interactions}

Although ultra-cold gases have extremely low densities (typically
below $10^{15}$~cm$^{-3}$), their properties are strongly influenced
by interactions~\cite{dalfovo,bec,pethick}. Usually, only the short
range, isotropic \emph{contact} interaction plays a role in quantum
degenerate gases.  However, another type of interaction, namely the
anisotropic, long range interaction between dipolar particles, has
attracted a lot of interest recently~\cite{baranov}.

\vspace*{-0.25cm}
\subsection{Contact potential}

In the ultra-low temperature regime relevant for quantum-degenerate
gases, scattering occurs only in the s-wave regime, as the
centrifugal barrier for other partial waves is much higher than the
typical kinetic energies of the atoms. A consequence of this fact is
that the real, complicated interatomic molecular potentials (which
at long distances are essentially given by Van der Waals attraction
$-C_6/r^6$) can be replaced for most purposes by a simple, isotropic
and short-range model potential proportional to the scattering
length $a$ of the atoms. This \emph{contact interaction} reads

\begin{equation}
U_{\rm contact}(\mathbf{r})=\frac{4\pi\hbar^2 a}{m}\delta(\mathbf{r})\equiv
g\delta({\mathbf{r}}),
\end{equation}
where $m$ is the atomic mass. Most of the interesting properties
observed in quantum gases since the first observation of BEC in dilute
atomic vapours in 1995 can be understood by using this simple
interatomic potential.

\vspace*{-0.25cm}
\subsection{Dipole dipole interaction (DDI) potential}
\label{sect_ddi}

\begin{figure}[t]
\includegraphics[width=12cm]{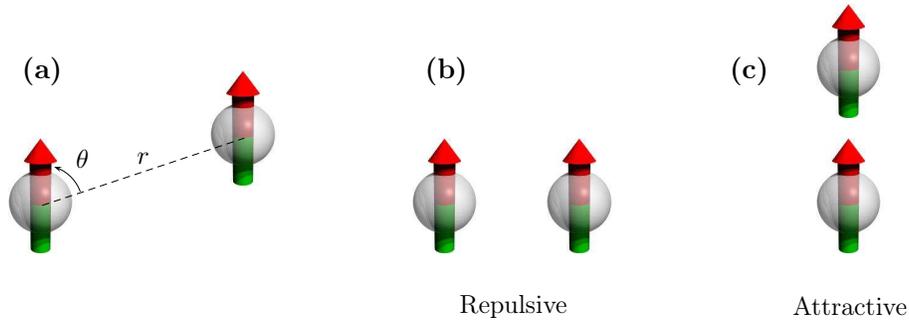}
\caption{(a) Notations for the dipole-dipole
interaction. (b) Dipoles placed side-to-side repel each other.  (c)
Dipoles in a `head-to-tail' configuration attract each other.}
\label{fig:attr}
\end{figure}

Atoms or molecules having a permanent dipole moment (either magnetic
or electric) interact not only via short-range potentials, but also
via the dipole-dipole interaction. The corresponding potential is

\begin{equation}
U_{\rm dd}(\mathbf{r})=\frac{C_{\rm
dd}}{4\pi}\frac{1-3\cos2\theta}{r^3}, \label{eq:udd}
\end{equation}
where $C_{\rm dd}$ is the dipolar coupling constant ($\cdd=\mu_0\mu^2$
for magnetic moments $\mu$, $\cdd=d^2/\varepsilon_0$ for electric
dipole moments $d$), and $\theta$ the angle between the direction
joining the two dipoles and the dipole orientation (we assume here
that all dipoles are aligned along the same direction~$z$). The DDI is
\emph{anisotropic} (dipoles
placed side-to-side repel each other, while dipoles in a head to tail
configuration attract each other, see Fig.\ref{fig:attr}) and \emph{long range}
(the $1/r^3$ dependence implies for example that the scattering
cross-section is not isotropic in the low-energy limit).

To characterise the relative strength of the dipolar and contact
interactions, it is convenient to introduce the dimensionless
parameter
\begin{equation}
\edd\equiv\frac{\cdd m}{12\pi\hbar^2 a}. \label{eq:edd}
\end{equation}
The numerical factors in $\edd$ are chosen such that a homogeneous BEC
with $\edd>1$ is unstable (see Sect.\ref{section:edd} below).  For the
alkalis usually used in BEC experiments, the value of $\edd$ is
extremely small (for example, for $^{87}$Rb, one has
$\edd\simeq0.007$), making the effects of the DDI negligible.

\subsection{Ultra-cold systems with long-range interaction}

There are several candidates to realize experimentally a dipolar
quantum gas: molecules having a permanent electric dipole moment $d$,
Rydberg atoms, which can have very large induced electric dipole
moments, or ground state atoms having a large magnetic moment $\mu$.
In this section, we are going to briefly describe the main
characteristics of those physical systems, and in the rest of this
review, we will describe in detail the case of ultra-cold Chromium
atoms, the only ones for which at present quantum degeneracy has been
achieved.

\subsubsection{Polar molecules}

Heteronuclear molecules in their ground state have a large electric
dipole moment, on the order of one Debye ($1\,{\rm D}\simeq 3.3 \times
10^{-30}\,{\rm C}\cdot{\rm m}$). Assuming that the order of magnitude
for the scattering length is similar to that of atoms used in BEC
experiments (typically around $100a_0$, where $a_0$ is the Bohr
radius), the corresponding value of $\edd$ is on the order of 100,
meaning that the properties of such a quantum gas would be dominated
by the DDI.

To date, no quantum degenerate gas of polar molecules is available
experimentally. Progress has been made recently in cooling of
molecules (see ref.~\cite{EPJD} for a review), but the densities and
temperatures achieved so far are still orders of magnitude away from
the quantum degenerate regime. A promising approach is to use Feshbach
resonances in mixtures of ultra-cold fermions in order to create
heteronuclear bosonic molecules.  Those molecules created in a highly
excited vibrational state must then be brought to the lowest
vibrational state, e.g. by photoassociation.

\subsubsection{Rydberg atoms}

Another system in which large electric dipole moments can be achieved
is given by Rydberg atoms. They are highly excited atoms, having large
principal quantum number $n$ and a dipole moment scaling like $n^2$.
The mutual interaction depends on the atomic states involved. By means
of an electric field, one can mix states with different electron
orbital angular momentum, such that the atoms acquire a dipole moment
and can interact to first order via dipole interaction.  Due to the
very large dipole moments, this interaction is very strong and is felt
over very long distances, of the order of many tens of microns.

The state of the art in this field includes the observation of the
blockade effect, which forbids more than one atom to be excited in a
given region of space (defined by the distance where the interaction
energy equals the linewidth of the excitation) and allows the
production of an atomic ensemble with a single collective excitation.
Moreover it includes the investigation of the resonant dipole-dipole
interaction in an electric field (F\"orster resonance), collective
behaviours and the real-time study of the dynamics of interacting
pairs of Rydberg atoms, revealing the character of the long-range
interactions.  For more details, see e.g.
\cite{rydberg} and references therein.

\subsubsection{Paramagnetic atoms with large magnetic moments $\mu$}

Some atoms like Cr, Eu, Dy, have a large magnetic moment of several
Bohr magnetons in their ground state. Only Cr has been Bose-condensed
to date, and therefore the only quantum gas to display measurable
dipolar effects is the Chromium BEC obtained in Stuttgart in
2004~\cite{jurgen}. Chromium has a magnetic dipole moment of $6\mub$,
and a scattering length of about $100 a_0$ ($a_0$ is the Bohr
radius). This gives $\edd\simeq0.16$, which allows to observe a
perturbative effect of the dipolar interaction.

\section{BEC of Chromium}

In this section, we give a very brief overview of the experimental
sequence used to obtain a $^{52}$Cr BEC. More details can be found
for example in~\cite{axel-phd}.

The specific level structure of Chromium makes it possible (and
necessary) to use novel laser cooling strategies to load atoms
continuously into a magnetic trap~\cite{clip}. After Doppler cooling
in the magnetic trap, we get a cloud of $1.5 \times 10^8$ atoms at a
temperature of a few hundreds of $\mu$K \cite{doppler}. RF-induced
evaporative cooling is then performed. However, dipolar relaxation
from the low-field-seeking state $\ket{^7{\rm S}_3,m_S=+3}$ towards
lower $m_S$ states prevents condensation \cite{dipolar-relax}.

The cloud (with $6 \times 10^6$ atoms at about 20~$\mu$K) is thus
transferred into a crossed optical dipole trap (50~W at 1070~nm), and
atoms are optically pumped to the high-field-seeking state
$\ket{^7{\rm S}_3,m_S=-3}$, which is the absolute ground state; with a
magnetic field of a few Gauss, dipolar relaxation is thus
energetically suppressed. Forced evaporative cooling in the dipole
trap then yields a pure condensate with up to $10^5$
atoms \cite{crbec}.

\subsection{Feshbach resonances in Chromium}

Besides its large magnetic moment, Cr has another asset: 14 Feshbach
resonances have been observed~\cite{joerg} for atoms in the state
$\ket{^7{\rm S}_3,m_S=-3}$. Close to such a resonance, the
scattering length varies with the applied magnetic field $B$ as
\begin{equation}
a(B)=a_{\rm bg}\left(1-\frac{\Delta}{B-B_0}\right).
\end{equation}
Here, $B_0$ is the resonance position, $\Delta$ its width, and
$a_{\rm bg}$ the background (non-resonant) scattering length. This
yields the possibility of increasing $\edd$ by making $a$ approach
zero. For this, one needs to control the magnetic field $B$ with a
precision much better than $\Delta$. The broadest known Feshbach
resonance of Cr lies at $B_0=589$~G, and has a width of
$\Delta=1.4$~G. The small value of
$\Delta/B_0\simeq2.3\times10^{-3}$ implies that one needs a good
control of the field  to tune $a$ accurately (typically, a relative control of the field at the $\sim 10^{-5}$ level is needed to control $a$ at the $\sim a_0$ level).

\subsection{Dipolar expansion}

The most spectacular effect of the magnetic dipole-dipole
interactions (MDDI) on the Cr BEC in an anisotropic harmonic trap appears in
time of flight experiments. The aspect ratio of the cloud during
expansion is modified by the MDDI and depends on the orientation of
the atomic dipoles with respect to the trap axes. In this section, we
first give a summary of the theoretical tools used to describe such
experiments, and then describe our results.

\subsubsection{GPE for dipolar gases}

\label{section:edd}

\paragraph{Pure contact interaction: a reminder}
Weakly interacting BECs with pure contact
interaction are well described by the Gross-Pitaevskii equation
(GPE) for the order parameter $\psi({\mathbf{r}},t)$

\begin{equation}
i\hbar\frac{\partial\psi}{\partial
t}=-\frac{\hbar^2}{2m}\triangle\psi+\left(V_{\rm
ext}+g|\psi|^2\right)\psi. \label{eq:gpe:cont}
\end{equation}
The non-linear term proportional to $g$ accounts for the effect of
interactions within the mean-field approximation. Note also that in
the time-independent case, the left-hand side of the above equation
has to be replaced by $\mu\psi$, with $\mu$ the chemical potential.
The normalisation of $\psi$ chosen here is $\int|\psi|^2=N$, where
$N$ is the total atom number.

A useful reformulation of the Gross-Pitaevskii equation is obtained
by writing $\psi=\sqrt{n}\exp(iS)$, with $n$ the atomic density and
$S$ the phase of the order parameter, related to the superfluid
velocity field by ${\mathbf{v}}=(\hbar/m){\nabla}S$. Substituting
in~(\ref{eq:gpe:cont}) and separating real and imaginary parts, one
gets the following set of hydrodynamic equations

\begin{equation}
\frac{\partial n}{\partial t}+{\nabla}\cdot(n{\mathbf{v}})=0,
\end{equation}
the equation of continuity, and an Euler-like equation
\begin{equation}
m\frac{\partial {\mathbf{v}}}{\partial t}%
+{\nabla}\left(\frac{mv^2}{2}+gn+V_{\rm
ext}-\frac{\hbar^2}{2m}\frac{\triangle
\sqrt{n}}{\sqrt{n}}\right)=0.
\end{equation}

For the case of a uniform condensate ($V_{\rm ext}=0$), one easily
shows, by linearising the hydrodynamic equations around equilibrium,
that the frequency $\omega$ and wavevector $k$ of a harmonic
perturbation are linked by the following dispersion relation, the
Bogoliubov spectrum

\begin{equation}
\omega=k\sqrt{\frac{gn}{m}+\frac{\hbar^2k^2}{4m^2}}.
\end{equation}

\paragraph{Dipolar interaction}

To include dipolar effects, one just needs (as long as the DDI
strength is not too high) to add an extra term to the mean-field
potential $g|\psi|^2$, namely

\begin{equation}
\fdd({\mathbf{r}},t)=\int|\psi({\mathbf{r}'},t)|^2\,U_{\rm dd}({\mathbf{r}}-{\mathbf{r}'})\,d {\bf r'}.
\end{equation}
This extra term is thus \emph{non-local} (due to the long-range
character of the DDI) and makes it much more complicated to solve the
GPE, even numerically (one faces now an integro-differential
equation).  For very large dipolar interaction (in the case, e.g., of
dipolar molecules), this simple modification of the Gross Pitaevskii
equation is not valid any more, as the separation of length scales on
which it relies (namely, that the scattering length entering the
contact interaction is determined by the intermolecular potential at
relatively small distances, where the effects of the DDI are small
compared to the van der Waals interaction) breaks down. In that case,
the contact interaction depends on the strength of the dipolar
coupling.

As a simple application, one can calculate the modifications to the
Bogoliubov spectrum induced by the DDI \cite{santosodell}. Using the
fact that the Fourier transform of the DDI~(\ref{eq:udd}) takes the
form

\begin{equation}
\widetilde{\udd}(\mathbf{k})=\cdd(\cos^2\alpha-1/3),
\end{equation}
where $\alpha$ is the angle between $\mathbf{k}$ and the direction of
the dipoles, and following the same method as in the previous
paragraph, one easily shows that the excitation spectrum is now given
by

\begin{equation}
\omega=k\sqrt{\frac{n}{m}\left[g+\frac{\cdd}{3}(3\cos^2\alpha-1)\right]+\frac{\hbar^2k^2}{4m^2}},
\end{equation}
and that, with the definition~(\ref{eq:edd}) for $\edd$, this implies
that a dipolar uniform condensate is unstable for $\edd>1$.

\subsubsection{Thomas-Fermi solutions and scaling Ansatz for the expansion}

\paragraph{Static Thomas-Fermi solutions}

For pure contact repulsive interaction, when the atom number is
increasing, the condensate size increases and the zero-point kinetic
energy becomes smaller and smaller. The Thomas-Fermi approximation
consists in neglecting the kinetic energy term in the time-independent
GPE; this gives then a simple algebraic equation, showing that the
density distribution has the shape of an inverted parabola.

\begin{figure}[t]
\includegraphics[width=13cm]{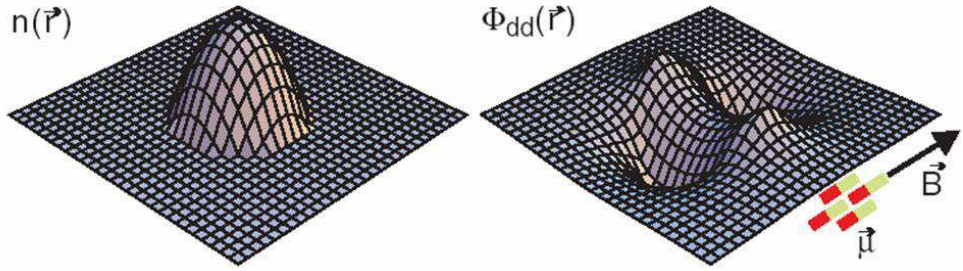}
\caption{ Density distribution for a
non-dipolar BEC in an isotropic trap (left). The resulting dipolar
potential $\fdd$ has a saddle-like shape, which tends to elongate the
condensate along the magnetization direction (right). Figure taken
from~\cite{jurgen}.} \label{fig:fdd}
\end{figure}

It is remarkable that this property remains valid if dipolar
interaction is included. This is due to the fact that the dipolar
mean-field potential $\fdd({\mathbf{r}})$ for a parabolic density
distribution $n({\mathbf{r}})=|\psi({\mathbf{r}})|^2$ is quadratic in
the coordinates (having a saddle shape because of the anisotropy,
Fig.\ref{fig:fdd})~\cite{stefano}.

For the case of a spherically symmetric trap (and thus also a
spherically symmetric density distribution for pure contact
interaction), one can easily show that, to first order in $\edd$, the
effect of the MDDI is to elongate the condensate along the direction
of magnetization: it is energetically favourable to accommodate new
particles close to the magnetization axis, where $\fdd(\mathbf{r})$ is
minimum (see Fig.\ref{fig:fdd}), thus causing an elongation of the
condensate. It is possible to show that this behaviour is valid for
anisotropic traps and for higher values of $\edd$. Note however that
for a non-spherical density distribution, calculating the coefficients
of the quadratic terms in $\fdd$ is possible but beyond the scope of
this paper. See~\cite{stefano} for details.

\paragraph{Expansion: scaling Ansatz}

For a pure contact interaction, and in the Thomas-Fermi approximation,
namely $Na/a_{\rm ho}\gg1$ with $a_{\rm
ho}=\sqrt{\hbar/(m\bar{\omega})}$, there exists a very useful solution
of the GPE~\cite{castindum}. It shows that the inverted parabola shape
of the condensate is maintained upon expansion (after release from the
trap), with a mere
\emph{rescaling} of its radii. The scaling parameters $b_i(t)$
($i=x,y,z$) giving the radii $R_i(t)=R_i(0)b_i(t)$ are solutions of
the ordinary differential equations

\begin{equation}
\ddot{b}_i=\frac{\omega_i^2(0)}{\displaystyle b_i\prod_{j\in
\{x,y,z\}} b_j}\qquad (i\in\{x,y,z\}).
\end{equation}

Like for the static case, this can be extended to the case of dipolar
interactions, since $\fdd$ keeps a parabolic shape. The corresponding
equations now read

\begin{equation}
\ddot{b}_i=\frac{\omega_i^2(0)}{\displaystyle b_i\prod_{j\in
\{x,y,z\}} b_j}+f(\{b_j\};\{\omega_j\};\edd)\qquad (i\in\{x,y,z\}),
\label{eq:stef}
\end{equation}
where $f$ is a function of the scaling radii, trap frequencies, and
dipolar parameter $\edd$. It turns out that the elongation of the
condensate along the magnetization direction remains valid during
expansion. The reader is again referred to~\cite{stefano} for further
details.

\subsection{Experiments}

\subsubsection{MDDI as a small perturbation}

\begin{figure}[t]
\includegraphics[width=8.25cm]{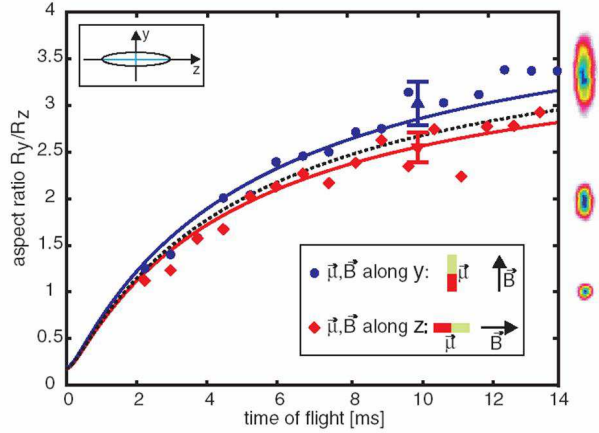}
\caption{ MDDI as a small perturbation in the
expansion of a condensate. The aspect ratio is measured during
expansion for two different orientations of the dipoles with respect
to the trap axes. Figure taken from~\cite{jurgen}.}
\label{fig:exp:pert}
\end{figure}

The first demonstration of an effect of the MDDI in a quantum gas
came soon after the first realization of a Cr BEC, by measuring the
aspect ratio of the BEC during time-of-flight for two different
orientations of the dipoles with respect to the trap axes. The small
value $\edd\simeq0.16$ implied that the effect was only a small
perturbation on top of the expansion driven by the contact
interaction (see Fig.~\ref{fig:exp:pert}).

\subsubsection{Use of the Feshbach resonance: strong dipolar regime}
\label{strong_dip}

To go beyond this perturbative effect, we used the 589~G Feshbach
resonance of Cr, in order to reduce $a$, and thus enhance $\edd$. We
provide this field using the offset coils of the magnetic trap, with
a current of about 400 amperes, actively stabilised at a level of
$4\times10^{-5}$ in relative value (peak to peak). The field is
switched on during the evaporation sequence in the ODT, at a stage
when the density is not too high, in order not to lose too many
atoms by inelastic losses when crossing the Feshbach resonances. The
rest of the experiment is performed in high field. After a BEC is
obtained, we ramp the field close to the resonance in 10~ms, hold
the field there for 2~ms, and take an absorption picture (still in
high field) after 5~ms of time of flight.

\begin{figure}[h]
{\bf(a)} \includegraphics[width=6.5cm]{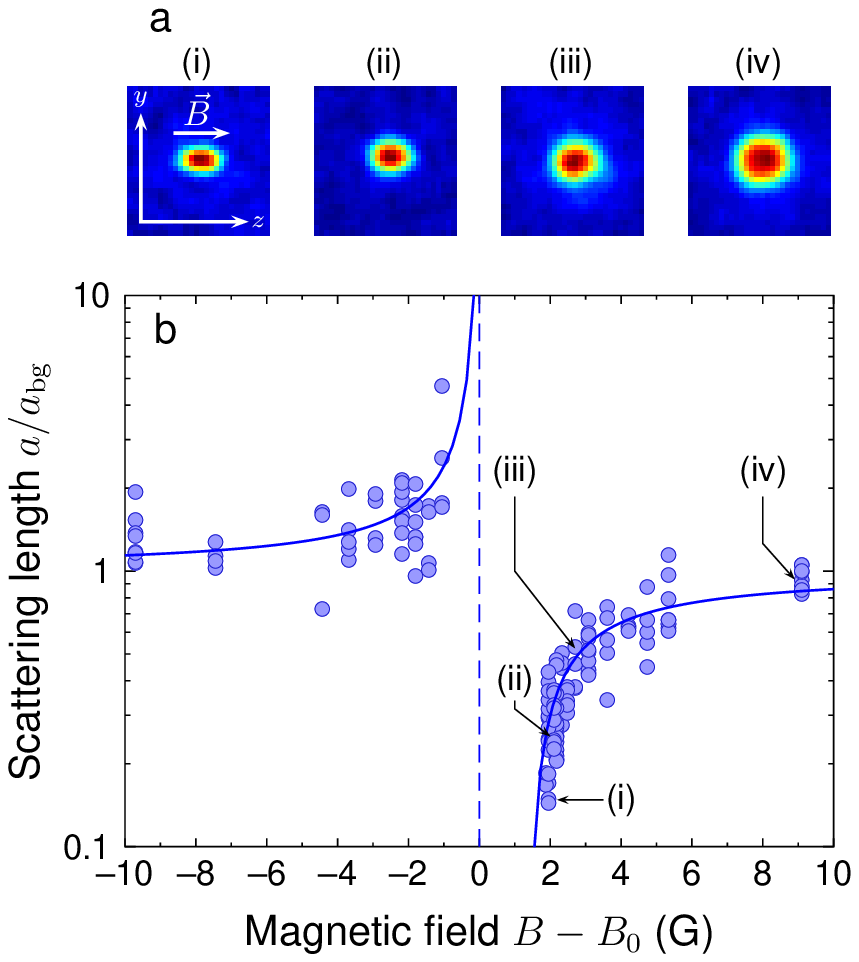} 
\qquad {\bf(b)} \includegraphics[width=6.25cm]{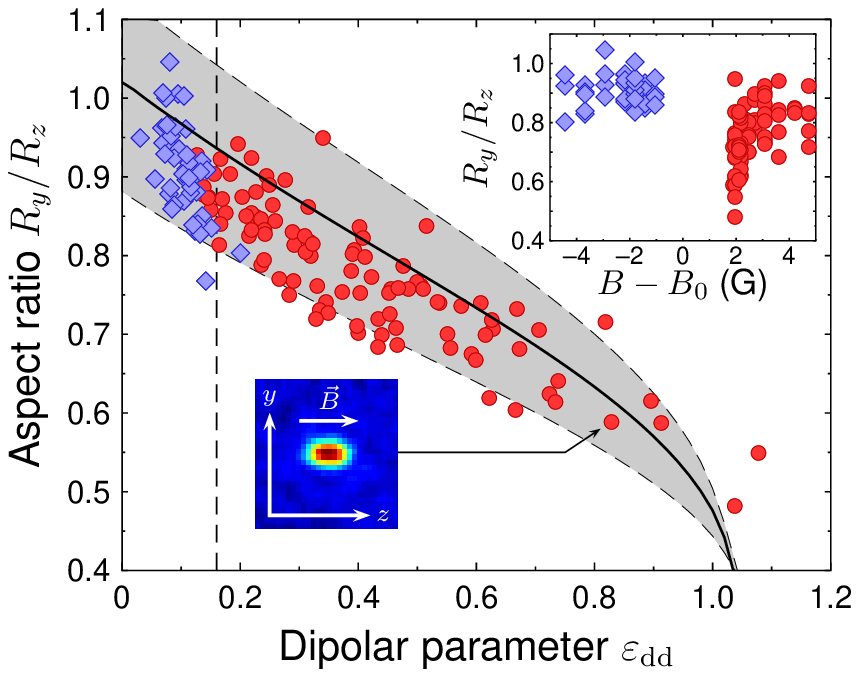} 
\caption{
(a) Measured scattering length $a(B)$ across the Feshbach resonance at
$B_0=589$~G. The condensate images (i)-(iv) in the top of the figure
clearly show a modification of the size (due to the reduction of $a$)
and of the aspect ratio (due to enhanced dipolar effects) of the
condensate.  (b) Aspect ratio of the BEC after 5~ms of expansion, as a
function of the measured dipolar parameter $\edd$. The solid line
corresponds to the prediction of hydrodynamic equations in the
Thomas-Fermi limit. Figure taken from~\cite{nature}.}
\label{fig:exp:ars}
\end{figure}

From the density distribution, we measure the Thomas-Fermi radii of
the BEC, and we infer the value of the scattering length (taking into
account explicitly the MDDI interaction by solving
Eqs.~(\ref{eq:stef})). The measured $a$ is shown on
Fig.~\ref{fig:exp:ars}~(a). One can see clearly a five-fold reduction
of $a$ above resonance, corresponding to a maximal value of
$\edd\simeq0.8$. On the sample absorption images of
Fig.~\ref{fig:exp:ars}~(a), one clearly sees, when $B$ approaches
$B_0+\Delta$, a strong reduction of the BEC size, due to the reduction
of $a$, and thus of the mean field energy released upon
expansion\footnote{For a pure contact interaction, the TF radius after
expansion scales as $(Na)^{1/5}$.}. But one also clearly observes an
elongation of the BEC along the magnetic field direction $z$. This
change in the cloud aspect ratio would not happen for a pure contact
interaction and is a direct signature of the MDDI.
Fig.~\ref{fig:exp:ars}~(b) shows the aspect ratio of the cloud as a
function of $\edd$, together with the theoretical prediction
from~(\ref{eq:stef}).

\begin{figure}[t]
\includegraphics[width=10cm]{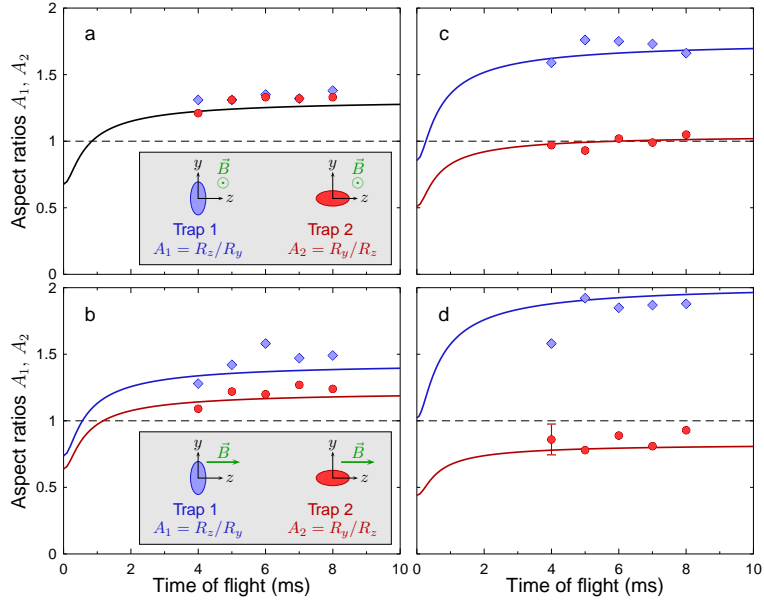}
\caption{ Aspect ratio of the BEC vs time of
flight. (a) $\edd=0.16$, magnetic field along $x$, the aspect ratio is
the same for the two trap configurations as expected. (b) $\edd=0.16$,
magnetic field along $z$, one basically recovers the results of
Fig.\ref{fig:exp:pert}. (c) $\edd=0.5$. d: $\edd=0.75$, the inversion
of ellipticity of the cloud is inhibited by the MDDI.  Figure taken
from~\cite{nature}.} \label{fig:exp:tofs}
\end{figure}

As an application of the tunability of $\edd$, we measured the aspect
ratio of the BEC during expansion for two different orientations of
the dipoles with respect to the trap axes. The effect of MDDI is now
way beyond the perturbative regime (Fig.\ref{fig:exp:tofs}). For large
enough $\edd$, one clearly sees that the usual inversion of
ellipticity of the BEC during expansion is inhibited by the MDDI.

Such tuning of the scattering length by means of a Feshbach resonance
is a first step in the study of the properties of purely dipolar
quantum gases ($\edd\gg 1$). Future directions of study include
e.g. the dependence of the stability of a dipolar BEC on the trap
geometry, or the behaviour of dipolar quantum gases in optical
lattices.

\section{Ultra-cold atoms in optical lattices}

The possibility of trapping and manipulating cold atomic gases with
the degree of freedom and precision described above allows to
investigate a huge range of different physical systems, where one has
control on the system geometry, the interatomic interactions and the
statistics of the atoms. One of the most fruitful fields of research
both from the experimental and the theoretical point of view in the
last years has been the study of ultra-cold atomic samples in optical
lattices.

Optical lattices are non dissipative periodic potential energy
surfaces for the atoms created by the intersection of laser
fields. The investigation of cold atoms in periodic potentials
allows to reproduce problems traditional of condensed matter and
solid state physics in a new setting, where a high degree of
control is possible and where the Hamiltonian which governs the
system is in general very close to some idealised one.

One of the greatest achievements of the last years was the
experimental observation of the superfluid to Mott insulator
transition (for details, see Sect.\ref{pointlike}) \cite{greiner,
fisher, jaksch}. In presence of long-range interactions, additional
quantum phases are predicted \cite{santos, sinha, sengupta, scarola},
like the supersolid phase, presenting the coexistence of superfluidity
and periodic spatial modulation, whose existence has not yet been
unambiguously proven experimentally \cite{helium}. The
search for this and other quantum phases makes cold atoms with long
range interactions particularly appealing.

As we will explain in the following, there are striking effects of
long range interactions to be expected for ultra-cold atoms in optical
lattices. Even if their existence strictly depends on the long-range
character of the interactions, their observability might require a
relative strength of the dipole-dipole interaction not too small
compared to the zero-range one. For this reason the recent
achievements with Chromium atoms presented in Sect.\ref{strong_dip}
are particularly important for the possible experimental realization
of the systems that we are going to discuss in the following.

First, we will introduce the main theory for ultra-cold atoms in
optical lattices in the extensively studied and experimentally already
realized case of point-like interaction. This will provide the
background for the topic of our interest, namely the case of
long-range interactions.

\subsection{Point-like interaction: superfluid to Mott insulator transition}
\label{pointlike}

As previously explained, in the most common cases, ultra-cold
atoms interact via s-wave scattering, which can be in a very good
approximation considered a point-like interaction.
Then, the Hamiltonian of the system in second quantisation
reads

\begin{eqnarray}
H&=&\int
\hat{\psi}^{\dag}({\bf r}) 
\left( \frac{p^2}{2m} + V_{\rm ext}({\bf r}) \right) 
\hat{\psi}({\bf r}) \;
d {\bf r} + \frac{g}{2} \int \hat{\psi}^{\dag}({\bf
r})\hat{\psi}^{\dag}({\bf r}) \hat{\psi}({\bf r}) \hat{\psi}({\bf r})
\; d {\bf r} \nonumber
\\ &&- \mu \int \hat{\psi}^{\dag}({\bf r}) \hat{\psi}({\bf
r}) \; d {\bf r} , 
\end{eqnarray}
where $ \hat{\psi}({\bf r})$ is the field operator, $V_{\rm ext}({\bf
r})$ the trapping potential, $g$ the interaction strength, linked to
the s-wave scattering length through $g=4\pi \hbar^2 a/m$, and $\mu$
the chemical potential, which fixes the total number of particles.

In the presence of an optical lattice, the trapping potential is
periodic $V_{\rm ext}({\bf r})=\sum_n V^{opt}_n sin^2(\pi x_n/d_n)$,
where the index $n$ runs over the dimensions of the lattice, and where
$d_n$ is the lattice constant in the $n$-th direction. For lattices
created by counterpropagating laser beams of wavelength $\lambda$, the
lattice spacing is $d=\lambda/2$. The intensity of the lattice
potential $V^{opt}_n$ depends on the intensity of the laser light.

As it is well known \cite{ashcroft}, the spectrum of a single
particle in a periodic potential is characterised by bands of
allowed energies and energy gaps. For deep enough periodic
potentials, the so-called tight binding regime is reached, where
the first band takes the form $E(q)=-2J \sum_n cos(q_n d)$, being
$q_n$ the quasi-momenta in the different lattice directions
and $J$ is the tunneling parameter between neighbouring wells.

Provided the lattice is deep enough and for low enough interactions
and temperature, the physics of the system can be well approximated by
the one taking place in the first energy band. The counterpart of the
energy eigenstates delocalised over the whole lattice (Bloch states)
are the wavefunctions localised at the bottom of each lattice site
(Wannier functions). In the tight binding regime is often convenient
to use the Wannier description.

In the single band approximation, the field operator can be
replaced by its single-mode expansion

\begin{eqnarray}
\hat{\psi}({\bf r})=\sum_i w_i({\bf r})\hat{a}_i , 
\end{eqnarray}
with ${\hat a}_i$ being the annihilation operator for one boson in the
Wannier function $w_i({\bf r})$ localised at the bottom of lattice
site $i$. Plugging this expression in the second quantised
Hamiltonian, one gets

\begin{eqnarray}
H &=& \sum_{i,j}
\int
 w_i^*({\bf r})
\left( \frac{p^2}{2m} + V_{\rm ext}({\bf r}) \right)
w_j({\bf r})\; d {\bf r} \;\;
\hat{a}_i^{\dag} \hat{a}_j +\\
&&+ \frac{g}{2} \sum_{i,j,l,m}
\int  w_i^*({\bf r}) w_j^*({\bf r}) w_l({\bf r})w_m({\bf r})
\; d {\bf r} \;\;
\hat{a}_i^{\dag} \hat{a}_j^{\dag}
\hat{a}_l \hat{a}_m
- \mu \sum_{i}  \hat{a}_i^{\dag} \hat{a}_i,
\nonumber
\end{eqnarray}
where in the chemical potential term we have exploited the
orthonormality of the Wannier functions.

For deep enough lattices, where the overlap beyond nearest
neighbours can be neglected, and defining

\begin{eqnarray}
J&=& - \int   w_i^*({\bf r})
\left( \frac{p^2}{2m} + V_{\rm ext}({\bf r}) \right) 
w_{i+1}({\bf r})  \; d {\bf r}  , \\
U&=&g \int  |w_i({\bf r})|^4 \; d {\bf r} ,
\end{eqnarray}
one gets, a part from constant terms,

\begin{eqnarray}
H= - J\sum_{\langle ij\rangle} \hat{a}_i^{\dag} \hat{a}_j +
\sum_{i}  \left[\frac{U}{2} n_i (n_i-1) - \mu n_i \right] ,
\label{bh}
\end{eqnarray}
with $n_i= \hat{a}_i^{\dag} \hat{a}_i$ being the density 
operator at site $i$.

This is the famous Bose-Hubbard Hamiltonian, extensively studied
in condensed matter physics. In optical lattices the Hamiltonian
parameters can be accurately controlled by changing the strength
of the optical lattice: ramping it up increases interactions due
to a stronger localisation of the wavefunctions at the bottom of
the lattice wells, and at the same time exponentially decreases
tunneling.

When tunneling is suppressed compared to interactions, this
Hamiltonian presents a quantum phase transition between a superfluid
phase, characterised by large number fluctuations at each lattice
site, and a Mott insulating phase where each lattice well is occupied
by precisely an integer number of atoms without any number
fluctuations. The nature of this phase transition and the qualitative
phase diagram can be inferred based on very simple arguments.

\begin{figure}
  \includegraphics[height=.12\textheight]{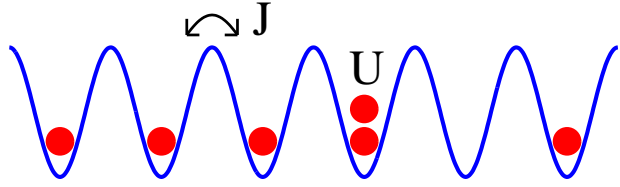}
\label{sf-mi}
\caption{Sketch of a particle-hole excitation in the $n=1$ Mott phase.
The energy cost for having two particles in the same site is $U$ and
the energy gained by hopping is $\sim J$. Their interplay determines
whether the ground state is insulating or superfluid.}
\end{figure}

At zero tunneling $J=0$ and commensurate filling (exactly an
integer number $n$ of atoms per well), the interaction energy is
minimised by populating each lattice well by exactly $n$ atoms.
Energy considerations can tell which is the range of
chemical potential $\mu$ at which the filling factor $n$ is the
most energetically convenient:

\begin{eqnarray}
&&E(n)=\frac{U}{2}n(n-1)-\mu n, \nonumber \\
&&E(n)<E(n+1)\;\;\;{\rm and}\;\;\; E(n)<E(n-1) \Rightarrow \nonumber \\
&&(n-1)U < \mu < n U . \label{boundary_o} \label{zeroJ}
\end{eqnarray}
This state with precisely integer occupation at the lattice sites is
called Mott insulating state.  Since a particle-hole excitation at
$J=0$ costs an energy $\Delta E=U$ equal to the interaction energy,
the Mott state is the lowest energy state at commensurate filling (see
Fig.\ref{sf-mi}). For a tunneling $J$ different from zero the energy
cost to create an excitation decreases thanks to the kinetic energy
favouring particle hopping. However, for large interactions and small
tunneling, the gain in kinetic energy ($\sim J$) is not yet sufficient
to overcome the cost in interaction energy ($\sim U$), which leads to
the existence of Mott insulating states also at finite tunneling. For
large enough tunneling, instead, particle hopping becomes
energetically favourable and the system becomes superfluid.
The regions in the $J$ vs. $\mu$ phase diagram where the Mott
insulating state is the ground state are called {\it Mott lobes}.
For non commensurate filling, there are extra atoms free to hop to
site to site at no energy cost, so that the phase of the system is
always superfluid. The superfluid phase at non commensurate densities
survives down to $J=0$ for $\mu/U=[\rho]$ where the symbol $[\rho]$
indicates the integer part of the density.

Due to the finite energy cost required to add or remove one
particle, the Mott phase is gapped and incompressible, while in
the superfluid regions the gap vanishes and the system is
compressible.

In Eq.(\ref{zeroJ}) we have identified the boundaries of the Mott
lobes at $J=0$. In order to find the shape of the lobes at finite $J$
more sophisticated calculations are required. In particular there is
no exact analytical method which allows to calculate them. A part from
the mean-field approximation \cite{fisher,sheshadri,sachdev}, which
is discussed later, high order perturbative strong coupling expansions
can be performed \cite{freericks}.  Exact numerical results can be
obtained using Quantum Monte Carlo techniques \cite{qmc}.

In the next sections, we introduce the mean-field approximation and
the perturbative method that we use to calculate the lobes. It works
only qualitatively in one dimension and work better and better in
larger dimensions.  This will provide further insight on the
conditions required for the insulating lobes to exist and proves very
useful to understand the analogies and differences in the case of
long-range interactions.

In the non uniform geometries available in the experiments, where a
harmonic potential is superimposed to the optical lattice, the
trapping potential is taken into account in the Bose-Hubbard
Hamiltonian through a site-dependent chemical potential
$\mu_i=\mu-V_i$, being $V_i$ the harmonic potential at site $i$. The
system is characterised by alternating shells of Mott and superfluid
phases \cite{jaksch, batrouni, bloch2}. In this paper we will
concentrate on the uniform system and not discuss the trapped system
in more detail.

\subsubsection{Mean-field decoupling and Gutzwiller Ansatz}

Starting from the Bose-Hubbard Hamiltonian (\ref{bh}), one can perform
the so-called mean-field decoupling \cite{fisher, sheshadri,
sachdev}. We define the order parameter $\langle a \rangle = \varphi$
and proceed with the following substitution

\begin{eqnarray}
a={\tilde a} + \varphi , \;\;
a^{\dag} = {\tilde a}^{\dag} + \varphi^*,
\end{eqnarray}
obtaining for the hopping term

\begin{eqnarray}
-J \sum_{\langle i j\rangle}\left[ {\varphi}^*_i \left( \hat{a}_j
- \frac{\varphi_j}{2} \right) + {\varphi}_j \left(
\hat{a}_i^{\dag} - \frac{\varphi_i^*}{2} \right)+  {\tilde
a}^{\dag}_i {\tilde a}_j  \right].
\end{eqnarray}
Based on this result, the full Hamiltonian can be rewritten with no
approximations as

\begin{eqnarray}
H= -J \sum_{\langle ij \rangle} \left[
 {\varphi}^*_i \left( \hat{a}_j - \frac{\varphi_j}{2} \right) +
{\varphi}_i \left( \hat{a}_i^{\dag} - \frac{\varphi_i^*}{2}
\right) + {\tilde a}^{\dag}_i {\tilde a}_j \right]+ \sum_i
\left[\frac{U}{2} n_i(n_i-1) - \mu n_i \right],
\end{eqnarray}
where the term ${\tilde a}^{\dag}_i {\tilde a}_j$ is supposed to be
small.

The full Hamiltonian in the mean-field approximation is just given by
the sum of the mean-field Hamiltonians on different sites
$H^{MF}=\sum_i H^{MF}_i$, where

\begin{eqnarray}
H^{MF}_i=
-J \left[
 {\bar \varphi}^*_i \left( \hat{a}_i - \frac{\varphi_i}{2} \right) +
{\bar \varphi}_i \left( \hat{a}_i^{\dag} - \frac{\varphi_i^*}{2} \right)
 \right]
+\frac{U}{2} n_i(n_i-1) - \mu n_i
\end{eqnarray}
and ${\bar \varphi}_i = \sum_{\langle j \rangle_i} \varphi_j$ takes
the neighbouring wells into account at the mean-field level.

In the homogeneous system, all order parameters $\varphi_i=\varphi$
are equal. For each set of parameters $U,J,\mu$, one can find the
value of $\varphi$ which minimises the ground state eingevalue (at
$T=0$). A vanishing order parameter is the indication of a Mott
insulating phase, while an order parameter different from zero, which
can be only produced by number fluctuations at the lattice sites, is
the signature of a superfluid state.
The mean-field solution found in this way is completely equivalent to
the one provided by the famous Gutzwiller Ansatz. This corresponds in
writing the state of the system as a product over the different
lattice sites of single-site wavefunctions

\begin{eqnarray}
|\Phi(t)\rangle= \prod_i \sum_n f_n^{(i)}(t) |i,n\rangle,
\end{eqnarray}
where $i$ is the site label and $n$ indicates the number state.
The coefficients $f_n^{(i)}$ are called Gutzwiller coefficients.

The solution obtained by the Gutzwiller Ansatz is intrinsically
mean-field. On the other hand it allows to extract many important
information and study the dynamics of the system. In fact, through a
time dependent variational approach based on the Lagrangian

\begin{eqnarray}
{\cal L}=
\frac{\langle{\dot \Phi}|\Phi\rangle - \langle\Phi|{\dot \Phi}\rangle}
{2i} - \langle \Phi|H|\Phi\rangle , 
\end{eqnarray}
one obtains the  dynamical equations for the Gutzwiller coefficients

\begin{eqnarray}
i \frac{d\,f_n^{(i)}}{dt} &=& -J \left[ \bar{\varphi}_i \sqrt{n_i}
  f_{n-1}^{(i)} + \bar{\varphi}_i^* \sqrt{n_i+1} f_{n+1}^{(i)} \right] +
 \left[\frac{U}{2} n_i(n_i-1)
  -\mu n_i \right] f_n^{(i)}.
\end{eqnarray}
By solving those equations in imaginary ($\tau=it$) time, one accesses
the stationary states of the system, while in real time the dynamics
of the system can be investigated also in the case of time-dependent
Hamiltonian parameters \cite{tdga}.

\subsubsection{Mean-field perturbative approach} \label{mfpa}

The mean-field perturbative approach allows to find the boundaries of
the lobes in an analytical way.  The main approximation underlying
this method is the mean-field approximation described in the previous
section.  For the purposes of this section, we separate the
Hamiltonian in two parts

\begin{eqnarray}
H_0&=& \sum_{i}  \left[ \frac{U}{2} n_i (n_i-1) - \mu n_i \right] ,\\
H_J&=&  - J \sum_i  \left[ {\bar \varphi}^*_i \left( \hat{a}_i -
\frac{\varphi_i}{2} \right) + {\bar \varphi}_i \left(
\hat{a}_i^{\dag} - \frac{\varphi_i^*}{2} \right) \right],
\end{eqnarray}
where we remind that $\varphi_i=\langle {\hat a}_i \rangle$ is the
order parameter and $\bar{\varphi_i}=\sum_{\langle j
\rangle_i}\varphi_j$ is the sum of the order parameters at sites
neighbouring to a given site $i$.

The term $H_J$ replaces the non-local tunneling term present in
the full Hamiltonian. Assuming that the order parameters
$\varphi_i$ are small, we rewrite $H_J$ at first order in the
order parameter as

\begin{eqnarray}
H_J \approx
-J \sum_i \left[ {\bar \varphi}_i^* \hat{a}_i +
{\bar \varphi}_i \hat{a}^{\dag}_i \right].
\end{eqnarray}
We use this approximate expression in the calculation of the
partition function

\begin{eqnarray}
{\cal Z}
&\approx& Tr \left[ e^{-\beta H_0}\right] - \int_0^{\beta} Tr
\left[ e^{-(\beta -\tau)H_0} H_J e^{-\tau H_0} \right] d\tau
+{\cal O}(\varphi^2) =   Tr \left[ e^{-\beta H_0}\right],
\end{eqnarray}
where we have used that $H_J$ is first order in the creation and
destruction operators and neglected second order terms. In the
limit $\beta \rightarrow \infty$, corresponding to zero
temperature, the partition function becomes

\begin{eqnarray}
{\cal Z} \equiv {\cal Z}_0 = e^{-\beta E_0},
\end{eqnarray}
being $E_0$ the energy of the ground state. In the calculation of
the order parameter instead, the first order term in the creation
and destruction operators also contributes, in the form

\begin{eqnarray}
\varphi_i
&\approx&
-  e^{\beta E_0} \int_0^{\beta}
Tr \left[ {\hat a}_i e^{-(\beta -\tau) H_0} H_J e^{-\tau H_0} \right] d\tau =
\nonumber \\
&=& J {\bar \varphi}_i e^{\beta E_0} \int_0^{\beta} Tr \left[ {\hat
a}_i e^{-(\beta -\tau) H_0} \hat{a}^{\dag}_i e^{-\tau H_0}
\right] d\tau , \label{trace}
\end{eqnarray}
where we have used the definition of $H_J$ at first order in the
order parameter, the fact that the expectation value of two
destruction operators over eigenstates of $H_0$ vanishes, and the
fact that non local averages vanish in the mean-field
approximation.
For the trace, we need to take into account all the matrix
elements of the kind
$\langle \Phi| {\hat a}_i e^{-(\beta -\tau) H_0} \hat{a}^{\dag}_i
e^{-\tau H_0} |\Phi \rangle$.
The two creation and annihilation operators in this expectation value
produce

\begin{eqnarray}
\left(n_i^{|\Phi\rangle}+1\right) exp\left[-\beta
H_0^{|\Phi+1,i\rangle}\right] exp \left[ -\tau \left(
H_0^{|\Phi\rangle} - H_0^{|\Phi+1,i\rangle}\right) \right],
\end{eqnarray}
where $|\Phi+1,i\rangle$ is the notation we use for the state
which is obtained from state $|\Phi\rangle$ by creating one more
particle at site $i$ and  $n_i^{|\Phi\rangle}$ is the number of
atoms at site $i$ in state $|\Phi\rangle$.

The integration over $\tau$ in the range $\tau \in [0,\beta]$, in
the limit $\beta \to \infty$ and the multiplication by $e^{\beta
E_0}$ give

\begin{eqnarray}
\left(n_i^{|\Phi\rangle}+1\right)
 \left[ \frac{exp\left[\beta
\left( E_0 - H_0^{|\Phi+1,i\rangle} \right)\right]} {
H_0^{|\Phi\rangle} - H_0^{|\Phi+1,i\rangle}} - \frac{
exp\left[\beta \left( E_0- H_0^{|\Phi\rangle} \right) \right]} {
H_0^{|\Phi\rangle} - H_0^{|\Phi+1,i\rangle}} \right].
\end{eqnarray}
Since by definition of ground state $E_0 - H_0^{|\Phi\rangle} \le 0$
for all states $|\Phi\rangle$, this result converges to a non zero
result if and only if either
\begin{eqnarray}
\label{condstab}
&1)& |\Phi\rangle \; {\rm is \; the \; ground \; state \; 
\; or} \\
&2)& |\Phi+1,i\rangle  \; {\rm is \; the \; ground \; state, \;
\; i.e. \;  
|\Phi\rangle \; is \; the \; state} \nonumber\\
&& {\rm obtained \; by \; removing \; one \; particle \; from \; the 
\; ground \; state \; at \; site}\; i. \nonumber
\end{eqnarray}
Namely, only two terms contribute to the trace (\ref{trace}): the one
arising from the ground state $\langle GS| {\hat a}_i e^{-(\beta
-\tau) H_0} \hat{a}^{\dag}_i e^{-\tau H_0} |GS\rangle$ and the one
arising from the states which are obtained from the ground state by
removing one particle $\langle GS-1,i| {\hat a}_i e^{-(\beta -\tau)
H_0} \hat{a}^{\dag}_i e^{-\tau H_0} |GS-1,i\rangle$.

Hence, for a Mott state with integer well occupation equal to $n_i$, the
order parameter reads

\begin{eqnarray}
\varphi_i = {\bar \varphi}_i J
\left[ \frac{n_i+1}{Un_i-\mu} + \frac{n_i}{\mu-U(n_i-1)}\right].
\label{M}
\end{eqnarray}
In a uniform system in the presence of point-like interaction
$\varphi_i$ is uniform and the solution can be found analytically.
Given that ${\bar \varphi}=z\varphi$, being $z$ the number of
nearest neighbours, the order parameter $\varphi=0$ unless

\begin{eqnarray}
1-zJ\left[ \frac{n+1}{Un-\mu} + \frac{n}{\mu-U(n-1)}\right]=0,\;\;
i.e.\;\; zJ=\frac{(Un-\mu)(\mu-U(n-1))}{\mu+U}. \label{lobes_pt}
\end{eqnarray}
This equation has solutions (for $zJ>0$) only for chemical potentials
in the range $n-1<\mu/U<n$, as already anticipated in
Eq.(\ref{zeroJ}).

Expression (\ref{lobes_pt}) of $\mu$ as a function of $J$ defines the
boundary of the Mott lobe at filling factor $n$. Outside of the lobes
the order parameter becomes different from zero, indicating the
arising of the superfluid phase. Then, this treatment, valid at 1st
order in $\varphi$, is not valid anymore to predict the value of the
order parameter.  However, in the framework of the mean-field
approximation, the expression for the lobes boundary is exact.
Moreover, most important, the prediction about the existence of the
lobes, i.e.  the lobes boundaries at $J=0$, are exact also beyond the
mean-field approximation and confirmed by exact numerical methods.

\begin{figure}[h]
  \includegraphics[width=7cm]{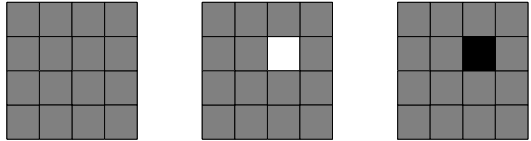}
\vspace*{0.5cm}
\label{ph}
\caption{Sketch of a Mott insulating $n=1$ phase with energy $E(n=1)$ 
(left); configuration where 1 atom has been removed, with energy
$E(n=1,-1,i)=E(n=1)+\mu$ (center); configuration where 1 atom has been
added, with energy $E(n=1,+1,i)=E(n=1)-\mu+U$ (right). For $0<\mu<U$,
the Mott $n=1$ phase is the ground state.}
\end{figure}

For the comparison that we will make later with the case of long-range
interaction it is useful to state the condition for the existence of
the lobes as follows: given an insulating state (classical
distribution of atoms in the lattice), this is stable in a given range
of chemical potential at $J=0$ if the energy increases by adding or
removing a particle at any site of the lattice, as depicted in
Fig.\ref{ph}.

\vspace*{0.5cm}

\subsubsection{Phase diagram}

\begin{figure}
  \includegraphics[height=.28\textheight]{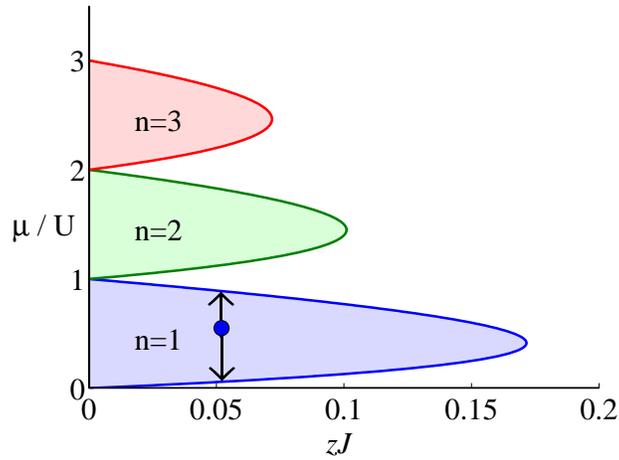}
\caption{Phase diagram in the mean-field approximation. The boundaries
of the Mott lobes are given by Eq.(\ref{lobes_pt}). The upwards and
downwards arrows describe respectively the energy of a particle and
hole excitation, as described in the text. Their sum is the energy of
a particle-hole excitation.}
\label{lobes} 
\end{figure}

Starting from Eq.(\ref{lobes_pt}) we can built the phase diagram of
the system in the mean-field approximation, as shown in
Fig.\ref{lobes}.  Inside the lobes the system in in the Mott
insulating phase. For a given tunneling parameter $J$ and chemical
potential $\mu$ the energy for adding or removing a particle from the
system is respectively given by the distance to the upper and lower
lobe boundaries (at constant $J$).  The sum of those energies (width
of the lobe at constant $J$) gives instead the energy for a
particle-hole excitation conserving the total number of atoms.

At the lobe boundary a quantum phase transition to the superfluid
phase takes place. The tip of the lobe is a special point, because
there the phase transition happens at commensurate filling factor $n$
and is purely due to the effect of increased tunneling.  At constant
non commensurate density the system is always superfluid.  For $J \to
0$, the lines of constant density accumulate between the Mott lobes,
at integer values of $\mu/U=n$ where the energies for having $n$ or
$n+1$ particle per site are equal.

\section{Long-range interactions in optical lattices}

\begin{figure}[h]
\includegraphics[height=.375\textheight]{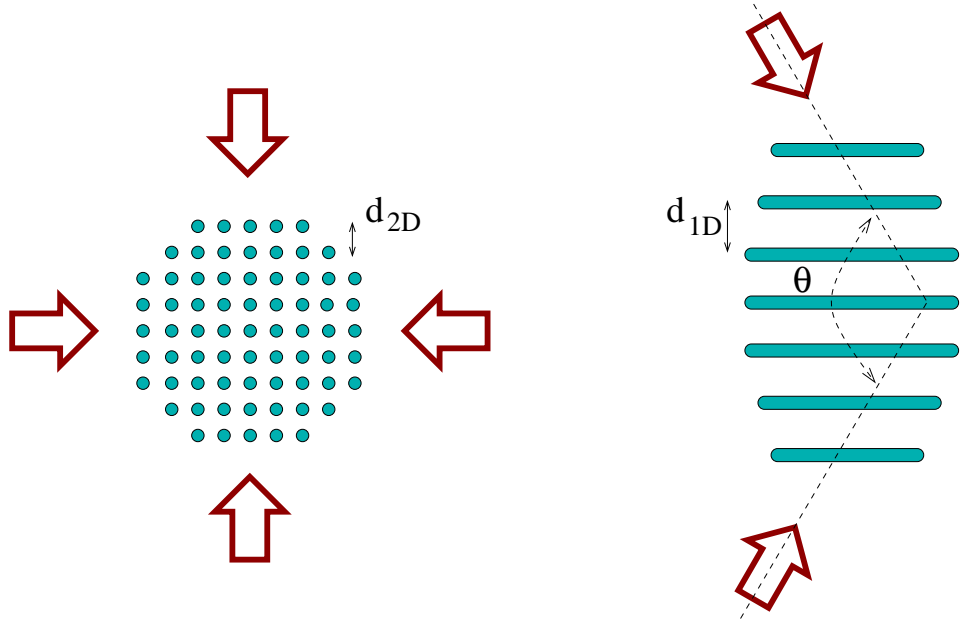}
\label{2D3D}
\caption{Two-dimensional lattice potential obtained by the intersection
of four counter-propagating laser beams. For a wavelength $\lambda$
the lattice spacing is given by $d_{2D}=\lambda/2$ (left); extra
one-dimensional lattice in the perpendicular direction, obtained by
the intersection of two laser beams at and angle $\theta$; the lattice
spacing is given by $d_{1D}=(\lambda/2)/sin(\theta/2)$ (right).}
\end{figure}

We now consider dipolar atoms in a 2D optical lattice.  In present
experiments, usually 2D geometries are created as a series of pancake
traps by means of a very strong 1D optical lattice in the
perpendicular direction, which provides strong confinement and
completely suppresses tunneling in one direction. In the presence of
long-range interaction, in order to consider each layer to be
isolated, one should not only suppress tunneling but also reduce the
interaction between the different layers.  For our purposes, where
each layer contains a 2D lattice, one should make the distance between
the different layers larger than the lattice spacing in the 2D
plane. This can be achieved by creating a 2D lattice with two pairs of
counterpropagating laser beams and the extra 1D lattice in the
perpendicular direction by two laser beams intersecting at a given
angle $\theta$, as depicted in Fig.\ref{2D3D}, in order to increase
the lattice spacing in the third direction to $d_{1D}=(\lambda/2)/
sin(\theta/2)$.

We consider an orientation of the dipoles perpendicular to the planes
of the 2D lattice.  This implies a repulsive interaction in the plane
and an attractive interaction between different layers. The effect of
this intra-layer attraction has been studied in \cite{demler}.  We
expect that the attractive character of the interaction in the
perpendicular direction might even help in reproducing in the
different layers the distribution of atoms that we find to exist in
the single plane. However, to draw exact conclusions one should devote
to this topic further studies.  In the following we are going to
restrict ourselves to the study of a single lattice plane.

\subsection{Extended Bose-Hubbard model}

In the presence of long-range interaction, an extra term has to be
added to the Hamiltonian of the system, which in second
quantisation reads

\begin{eqnarray}
H&=&\int
\hat{\psi}^{\dag}({\bf r})
\left( \frac{p^2}{2m} + V_{\rm ext}({\bf r}) \right) \hat{\psi}({\bf r})
\; d {\bf r}
+ \frac{g}{2} \int \hat{\psi}^{\dag}({\bf r})\hat{\psi}^{\dag}({\bf r})
 \hat{\psi}({\bf r})  \hat{\psi}({\bf r}) \; d {\bf r} + \nonumber \\
&+& \int\int \hat{\psi}^{\dag}( {\bf r'})\hat{\psi}^{\dag}( {\bf r'}) 
U_{\rm dd}( {\bf r}- {\bf r'})
 \hat{\psi}({\bf r})  \hat{\psi}({\bf r}) \; d {\bf r} \; d {\bf r'}
- \mu \int \hat{\psi}^{\dag}({\bf r})  \hat{\psi}({\bf r}) d  {\bf r} ,
\end{eqnarray}
where $U_{\rm dd}({\bf r})$ is the dipole-dipole potential
defined in Eq.(\ref{eq:udd}). 

\begin{figure}[h]
\includegraphics[scale=1]{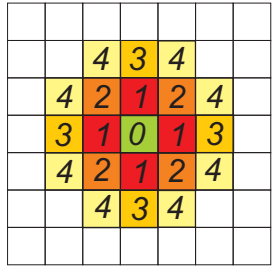}
\caption{Representation of the first four nearest neighbours in a 
2D optical lattice.} 
\label{NN}
\end{figure}

In the single band approximation, the generalised Bose-Hubbard
Hamiltonian in the presence of long-range interaction becomes

\begin{eqnarray}
H&=&
-J\sum_{\langle ij\rangle} \hat{a}_i^{\dag} \hat{a}_j
+ \sum_{i} \left[\frac{U}{2} n_i (n_i-1) - \mu n_i  \right]
+ \sum_{\vec \ell} \sum_{\langle i j \rangle_{\vec \ell}}
\frac{U_{\vec \ell}}{2} \; n_i n_j ,
\end{eqnarray}
where ${\vec \ell}$ is the distance connecting the two optical lattice
sites $i$ and $j$. The sum over the distance ${\vec \ell}$ is cut-off
at a certain nearest neighbour. In our calculations we usually
considered up to the $4th$ nearest neighbour, as shown in
Fig.\ref{NN}. Longer interaction ranges have a crucial effect on the
regions of the phase diagram describing low density of particles and
holes \cite{trefzger}, but we are not going to discuss this here.

\vspace*{0.5cm}
\vspace*{0.5cm}

The on-site interaction parameter $U$ is now given by two
contributions: one is, as before, arising from the $s$-wave scattering
$U_{\rm s}= 4 \pi \hbar^2 a/m \int n^2(r) d^3 r$, and the second one
is due to the on-site dipole-dipole interaction $U_{\rm dip}= 1/(2\pi)
\int {\widetilde U_{\rm dd}}(q) {\tilde n}^2(q) d^3q$, 
being ${\widetilde U_{\rm dd}}(q)$ and ${\tilde n}(q)$ the Fourier
transform of the dipole potential and density, respectively
\cite{goral}.

Because of the localisation of the wavefunctions at the bottom of the
optical lattice wells, the long range part of the dipole-dipole
interaction $U_{\vec \ell}$ is in a very good approximation given by
the dipole-dipole interaction potential at distance ${\vec \ell}$,
$U_{\vec \ell}= (C_{\rm dd}/4 \pi) [1-3\cos^2(\theta_{\vec
\ell})]/\ell^3$ 
multiplied by the densities $n_i$ and $n_j$ in the two sites, where
the quantity $C_{\rm dd}$, proportional to the dipole moment squared,
has been defined Sect.\ref{sect_ddi} and $\theta_{\vec \ell}$ is the
angle between ${\vec \ell}$ and the orientation of the dipoles
(see Fig.\ref{fig:attr}).

\begin{figure}[h]
  \includegraphics[height=.4\textheight,angle=270]{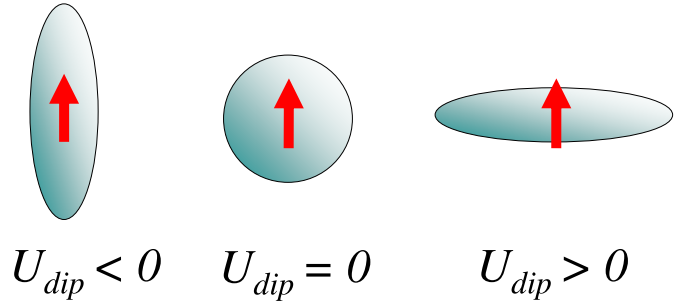}
\vspace*{0.5cm}
\caption{On-site dipole-dipole interaction $U_{\rm dip}$ depending on the
anisotropy of the wavefunction at the bottom of the lattice wells.
For vertically pointing dipoles, the dipole-dipole interaction is
mainly attractive for cigar-shaped wells (left), vanishes for
spherical wells (center) and is mainly repulsive for pancake-shaped
wells (right).}
\label{anisotropy} 
\end{figure}

The ratio between the total on-site interaction $U=U_{\rm s}+U_{\rm
dip}$ and the nearest neighbour dipolar interaction $U_{NN}$ determines
much of the physics of the system. As we mentioned previously, in
order to make the effects of long-range interactions observable it
might be necessary to have a not too small ratio $U/U_{NN}$. It can be
varied by tuning the strength and the sign of the on-site
dipole-dipole interaction $U_{\rm dip}$ by changing the anisotropy of
the Wannier functions at the bottom of the lattice sites, as depicted
in Fig.\ref{anisotropy}.  In standard experiments with $^{52}$Cr,
$U/U_{NN}
\approx 400$ (for $\varepsilon_{\rm dd}= \mu_0 (6\mu_B)^2 m/(12 \pi
\hbar^2 a) \approx 0.158$ and spherical localisation at the bottom of
the potential well at $s=20E_R$, where $E_R$ is the recoil energy at
$\lambda=500 \;{\rm nm}$). Using a Feshbach resonance to change the
$s$-wave scattering length, as recently demonstrated with Chromium
atoms \cite{nature}, $U/U_{NN}$ can be virtually tuned down to zero.

\vspace*{0.5cm}
\vspace*{0.5cm}

Analogously as before, a time dependent variational principle leads to
the dynamical equations for the Gutzwiller coefficients which now
include a contribution due to the long-range part of the interaction

\begin{eqnarray}
i \frac{d\,f_n^{(i)}}{dt} &=& -J \left[ \bar{\varphi}_i \sqrt{n_i}
  f_{n-1}^{(i)} + \bar{\varphi}_i^* \sqrt{n_i+1} f_{n+1}^{(i)} \right] +   \\
&+& \left[\frac{U}{2} n_i(n_i-1) + \sum_{\vec \ell} U_{\vec \ell}\;
  \bar{n}_{i,{\vec \ell}} \; n_i -\mu n_i \right] f_n^{(i)}, \nonumber
\end{eqnarray}
with $\bar{n}_{i,{\vec \ell}} = \sum_{\langle j \rangle_{i,{\vec
\ell}}} n_j $.

The imaginary time evolution, which mimics dissipation in the
system, converges unambiguously to the ground state of the system
for the Bose-Hubbard model in presence of on-site interaction only.
In the presence of long-range interaction it shows a strikingly
different behaviour and converges often to different
configurations, depending on the exact initial conditions. In this
way, we clearly get a feeling of the existence of metastable
states in the system. In the real time evolution, their stability
is confirmed by typical small oscillations around a local minimum
of the energy.

\subsubsection{Conditions to have an insulating lobe}

The existence of metastable stationary states can be confirmed using
the mean-field perturbative approach introduced in Sect.\ref{mfpa}.
In this section we will generalise it to include long-range
interaction and we show that, beyond the ground state lobe, 
insulating lobes for the metastable states exist \cite{menotti}.

The first important difference is that in the limit $\beta \rightarrow
\infty$, the system can populate any local minimum of the energy
and not necessarily the ground state, as it was before. Hence,
 the partition function becomes

\begin{eqnarray}
{\cal Z} \equiv {\cal Z}_{MS} = e^{-\beta E_{MS}},
\end{eqnarray}
where $MS$ indicates any of the metastable states, whose existence
we are for the moment assuming and going to demonstrate in the
following.

The expectation value of the order parameter reads, similarly to
before

\begin{eqnarray}
\varphi_i&=&
J  {\bar \varphi}_i e^{\beta E_{MS}} \int_0^{\beta}
Tr \left[ {\hat a}_i e^{-(\beta -\tau) H_0} \hat{a}^{\dag}_i
e^{-\tau H_0} \right] d\tau , \label{trace2}
\end{eqnarray}
where $H_0$ now includes also the dipole-dipole interaction

\begin{eqnarray}
H_{0}=\sum_i \left[ \frac{U}{2} n_i(n_i-1) - \mu n_i + \sum_{\vec
\ell} U_{\vec \ell}\; {\bar n}_i \left( n_i -\frac{\langle n_i
\rangle}{2}\right) \right].
\end{eqnarray}
On the other hand, the tunneling part $H_J$ has not changed
with respect to the case of zero-range interaction.

It is important to stress at that point that in order to investigate
the stability of a given metastable state, the trace is performed on
the subspace of states differing from it only by small
perturbations. Specifically this means that we take into account only
those states which differ from the metastable state by adding or
removing one particle.

The derivation follows the one presented in Sect.\ref{mfpa}. The
important analogies and differences are summarised here:

a) the ground state is replaced by any metastable state and only small
perturbations around that metastable state are considered;

b) the assumption of metastability of a given configuration has to
ensured by checking that all the states obtained from the metastable
state $|MS\rangle$ by adding or removing one particle at {\it any} of
the lattice sites $j$ have higher energy than $E_{MS}$, i.e.

\begin{eqnarray}
E_{MS} - H_0^{|\Phi\rangle} \le 0, \hspace*{0.5cm} 
{\rm for}\;\; |\Phi\rangle = |MS \pm 1, j\rangle\;\; \forall j;
\label{b}
\end{eqnarray}

c) the only non vanishing contributions arise, analogously to
Eq.(\ref{condstab}), from the terms where

\begin{eqnarray}
\label{condstabMS}
&1)& |\Phi\rangle  \; {\rm is \; the \; metastable \; state \; 
\; or} \\
&2)& |\Phi+1,i\rangle  \; {\rm is \; the \; metastable \; state, \;
\; i.e. \;  
|\Phi\rangle \; is \; the \; state} \nonumber\\
&& {\rm obtained \; by \; removing \; one \; particle \; from \; the 
\; metastable \; state \; at \; site}\; i. \nonumber
\end{eqnarray}
These are exactly the same criteria as for on-site interaction only, a
crucial difference being that now the atomic distribution in the
lattice sites might be not uniform so that the effect of adding and
removing one particle at {\it all} lattice sites has to be considered
to ensure condition (\ref{b}) above.
Taking the case of dipole-dipole interaction explicitly into account,
one gets that for all $i$ the conditions

\begin{eqnarray}
U(n_i-1)<\mu-V^{1,i}_{dip}<Un_i \hspace*{0.5cm} 
\end{eqnarray}
have to be satisfied, implying for the boundary of the lobes
at $J=0$ the values

\begin{eqnarray}
&&\mu_{min}=\max_i\left[U(n_i-1)+V^{1,i}_{dip}\right] , \\
&&\mu_{max}=\min_i\left[Un_i+V^{1,i}_{dip}\right] ,
\end{eqnarray}
where $V^{1,i}_{dip}$ is the dipole-dipole interaction of 1 atoms at
site $i$ with the rest of the lattice.
Clearly in the absence of dipole-dipole interaction, these
conditions reduce to the ones written in Eq.(\ref{boundary_o}).

\begin{figure}[h]
  \includegraphics[width=7cm]{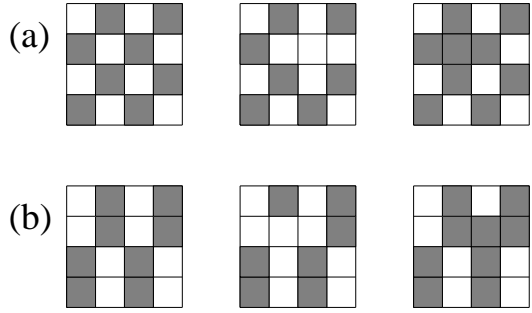}
\label{m_ph}
\caption{(a) Sketch of a checkerboard Mott insulating phase with energy 
$E(GS)$ (left); configuration where 1 atom has been removed with
energy $E(GS,-1,i)=E(n=1)+\mu$ (center); configuration where 1 atom
has been added with energy $E(n=1,+1,i)=E(n=1)-\mu+4U_{NN}$
(right). For $0<\mu<4U_{NN}$, the Mott $n=1$ phase is the ground state.
(b) Sketch of a metastable Mott insulating phase at filling factor
$1/2$ with energy $E(MS)$ (left); configuration where 1 atom has been
removed with energy $E(MS,-1,i)=E(n=1)+\mu-U_{NN}$ (center);
configuration where 1 atom has been added with energy
$E(n=1,+1,i)=E(n=1)-\mu+3U_{NN}$ (right). For $U_{NN}<\mu<3U_{NN}$,
this metastable state is stable.  For the sake of simplicity, in those
examples we have considered only one nearest neighbour in the
dipole-dipole interaction.}
\end{figure}

Those stability conditions immediately lead us to draw two important
conclusions:

i) configurations with no integer filling factor can be stable as soon
as one nearest neighbour is included in the interaction (see
Fig.\ref{m_ph}(a));

ii) for a given chemical potential there exist configurations at
higher energy compared to the ground state which also fulfill the
stability condition (see Fig.\ref{m_ph}(b)). We call such
configurations {\it metastable states}.

The explicit expression for the order parameter now reads

\begin{eqnarray}
\varphi_i 
&=& \bar{\varphi_i} J
\left[\frac{n_i+1}{Un_i-\mu+V^{1,i}_{dip}}-
\frac{n_i}{U(n_i-1)-\mu+V^{1,i}_{dip}}\right].
\label{M2}
\end{eqnarray}
Equation (\ref{M2}) is a system of linear equations in the
variables $\varphi_i$ described by the matrix M: $M {\vec \varphi}=0$.
When $det(M)\neq 0$, this equation only allows the trivial solution
$\varphi_i=0$ $\forall i$, implying that the system is in a Mott
insulating phase. On the other hand, if $det(M)=0$, it is possible to
have $\varphi_i \neq 0$, implying that the system has become superfluid.

Due to the non homogeneity of the solution in the presence of
long-range interaction this is a system of linear equations whose
dimension depends on the size of the lattice. In practise we can
access with this method lattice sizes of the order of 20-30 lattice
sites per direction.

\subsubsection{Phase diagram for long-range interactions}

\begin{figure}
  \includegraphics[height=.5\textheight]{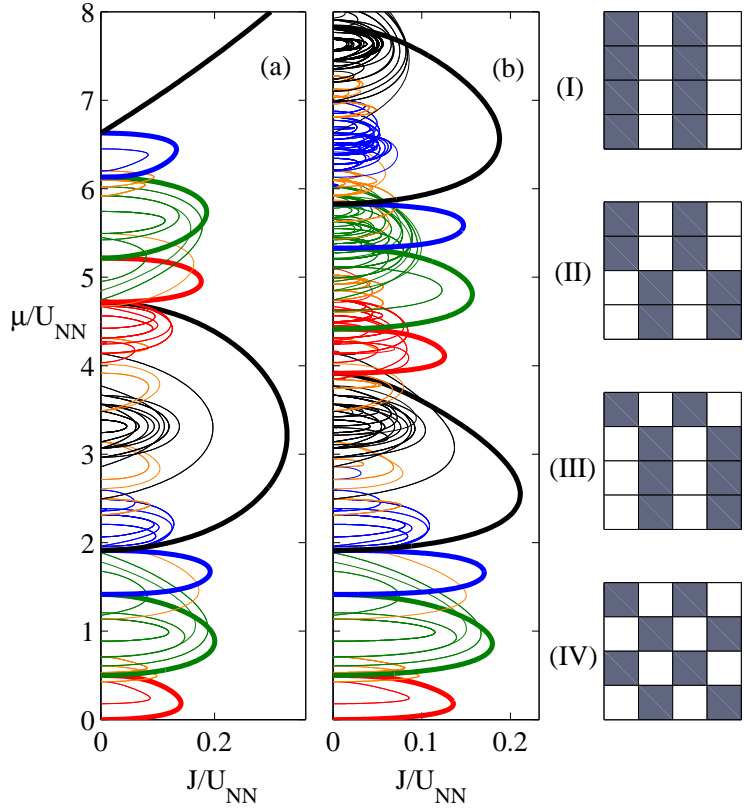}
\label{lobes_dipdip}
\caption{ Phase diagram for weak and strong dipole-dipole interaction
and interaction range up to the 4th nearest neighbour: $U/U_{NN}=20$
(a) and $U/U_{NN}=2$ (b). The thick lines are the ground state lobes,
found (for increasing chemicals potential) for filling factors equal
to all multiples of 1/8. The thin lines are the metastable states,
found at filling factors equal to multiples of $1/16$. Some of the
metastable configurations at filling factor 1/2 (I to III) and
corresponding ground state (IV).  Empty sites are light and sites
occupied with 1 atom are dark. Figure from \cite{menotti}.}
\end{figure}

The phase diagram of the system is shown in Fig.\ref{lobes_dipdip}.
As predicted by the perturbative mean-field approach presented in the
previous section, in the presence of long-range interactions, there
are insulating lobes corresponding both to ground and metastable
states and to integer and non integer filling factors. In particular 
non uniform distributions of the atoms are allowed.

The phase diagram in Fig.\ref{lobes_dipdip} has been calculated for an
interaction range of four nearest neighbours, as shown in
Fig.\ref{NN}. For this specific range of interaction, the lowest
filling factor allowed is $1/8$.

By increasing the relative strength the long-range interaction with
respect to the on-site one, the lobes corresponding to fractional
filling factors become comparable to the ones relative of the standard
Mott phase with commensurate filling (cfr.  Fig.\ref{lobes_dipdip}(b)
with Fig.\ref{lobes_dipdip}(a), where the upper lobe for $n=1$ is only
partially shown).

This phase diagram is confirmed by the imaginary and real time
evolution of the system: depending on the initial conditions the
imaginary time evolution can converge to the metastable
configurations, while in the real time evolution, their stability is
reflected into typical small oscillations around a local minimum of
the energy. In the next section we discuss the lifetime of the
metastable states, connected to the possibility of tunneling between
different local minima of the energy landscape.

\subsection{Stability of the metastable states}

\begin{figure}[h]
\includegraphics[width=4cm,angle=0]{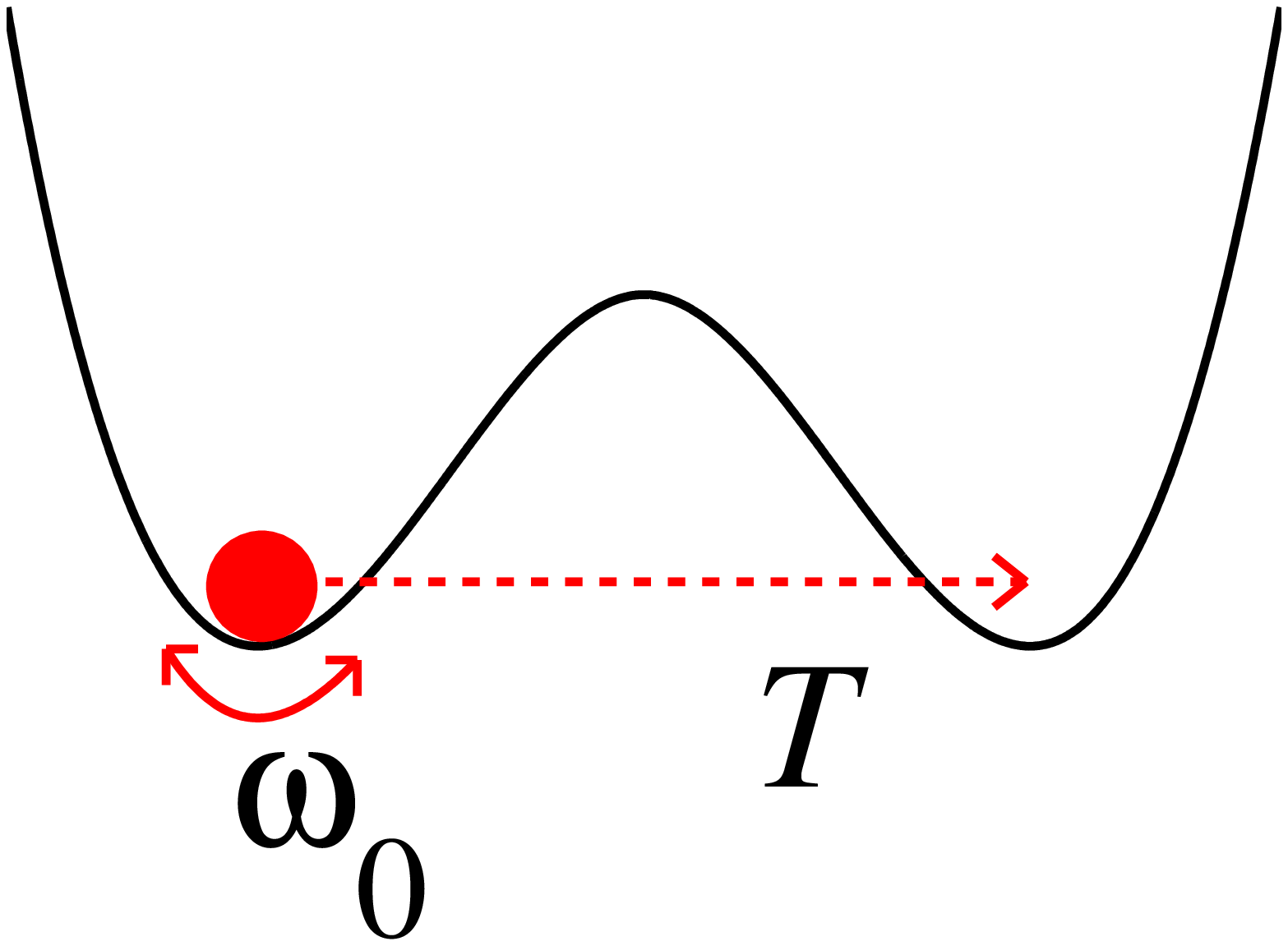} \hspace{3cm}
\includegraphics[width=4cm,angle=0]{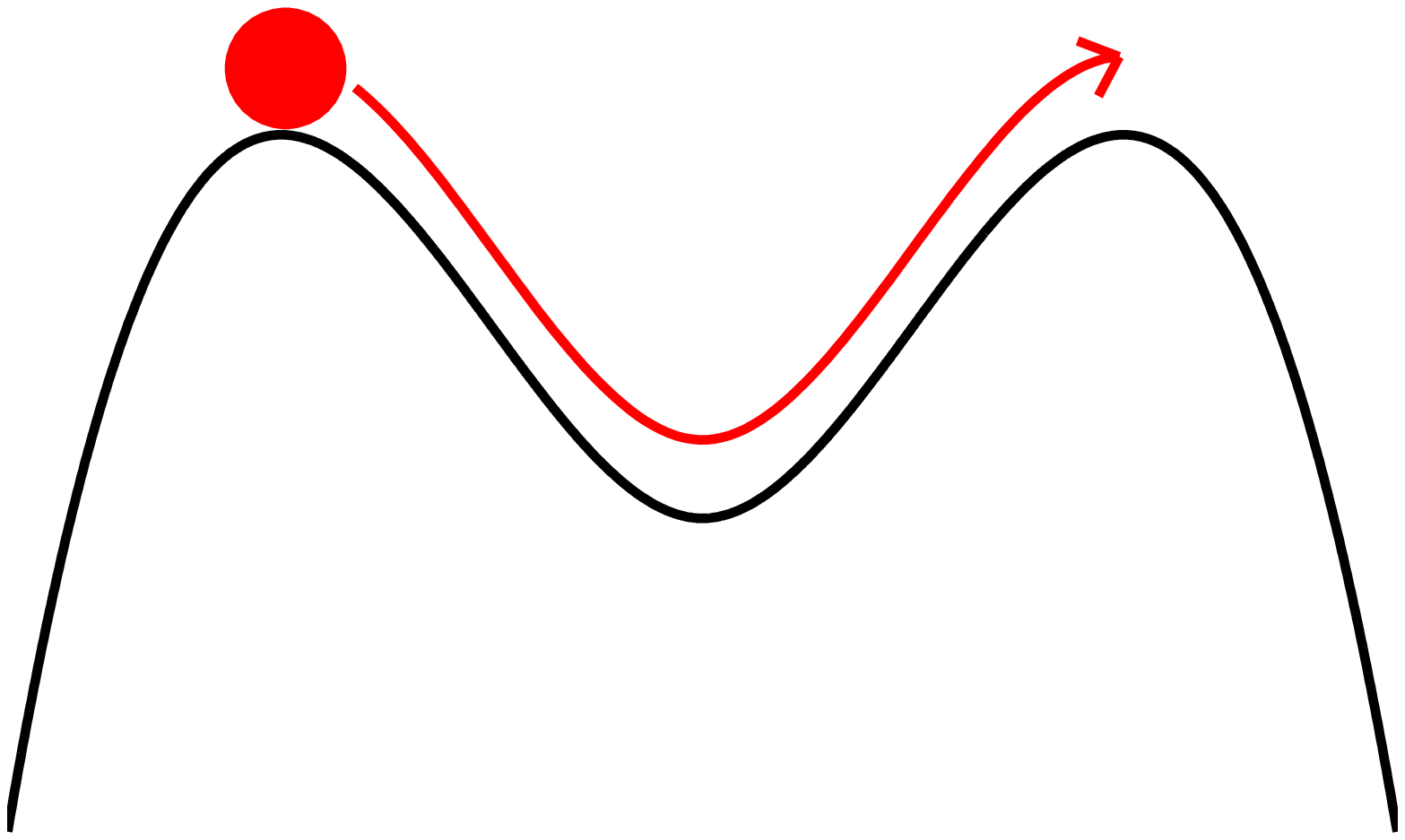}
\caption{Tunneling of a particle in a double-well potential (left);
in imaginary time, this tunneling event (instanton) is described by
the dynamics in the inverted potential (right).}
\label{inverted_pot}
\end{figure}

We study the stability of the metastable states with a path integral
formulation in imaginary time \cite{wen}, which can describe the
tunneling below a potential barrier (instanton effect).  The path
integral representation for the propagator in imaginary time takes the
form

\begin{eqnarray}
\int {\cal D}[\Phi^*,\Phi] \prod_{j=1}^N \langle
\Phi_j|e^{-H\Delta \tau}|\Phi_{j-1}\rangle \approx 
\int {\cal D}[\Phi^*,\Phi]  e^{-\int {\cal L}(\tau,\Phi,{\dot \Phi}) d\tau},
\end{eqnarray}
which gives the following expression for the Lagrangian ${\cal
L}=\langle \Phi|{\dot \Phi}\rangle - H$.  With the substitution
$\Phi=(x+ip)/\sqrt{2}$ and $\Phi^*=(x-ip)/\sqrt{2}$, the Lagrangian
becomes ${\cal L}=-ip{\dot x}+H$. This can be put in the canonical
form ${\cal L}=P{\dot X}-{\tilde H}(X,P)$ by defining the new
coordinates $X=x$ and $P=\partial {\cal L}/\partial {\dot X}=-ip$. In
those new coordinates, the Hamiltonian governing the dynamics in
imaginary time is

\begin{eqnarray}
{\tilde H}(X,P)=-H(p(P),X)=\frac{P^2}{2m}-V(X).
\end{eqnarray}
For this reason, the dynamics in imaginary time, describing the
instanton effect between two potential wells, is said to happen in
the inverted potential $-V(X)$, as depicted in Fig.\ref{inverted_pot}.

\vspace*{0.25cm}

We exploit this formalisms to describe the tunneling between the
different metastable configurations found in the previous section.  In
order to reduce our problem to the dynamics of a single variable $X$
and its conjugate momentum $P$, we make a variational Ansatz on the
Gutzwiller wavefunction which parametrises the population of the
lattice well from 1 to 0 and viceversa.

\vspace*{0.25cm}
\begin{figure}[h]
\includegraphics[width=5cm,angle=0]{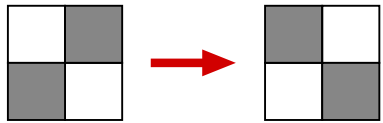}
\caption{Transition from a checkerboard configuration to
its opposite one, where full sites are replaced by empty sites and
viceversa.} 
\label{transition}
\end{figure}

\begin{figure}[h]
\includegraphics[width=7cm,angle=0]{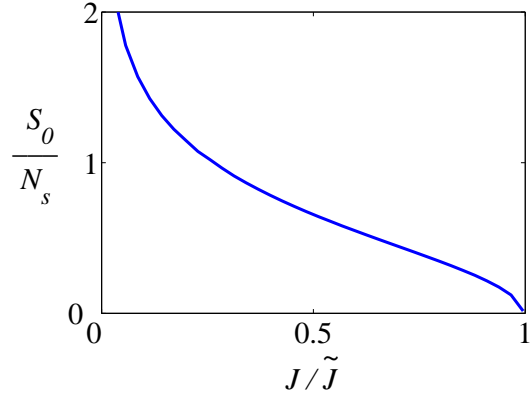}
\caption{Action $S_0$ for the transition described in Fig.\ref{transition},
calculated with a variational Ansatz for the path integral formalisms
in imaginary time. The lifetime of the tunneling event is given by
$\omega_0 T \approx \exp[S_0]$. $N_s$ is the number of sites which
invert their population in the transition and ${\tilde J}$ is of the
order of the tip of the insulating lobe.}
\label{action}
\end{figure}

Here we present in particular the results for the transition between
one configuration (in the specific case, the checkerboard) and its
opposite one, where full and empty wells have been interchanged, as
shown in Fig.\ref{transition}.
For this transition, the tunneling time is $\omega_0 T \approx
\exp[S_0]$, where $\omega_0$ is the typical oscillation frequency
around the minimum of the energy. The tunneling time diverges for $J
\to 0$ and roughly scales like $\omega_0 T \approx \exp[N_s\hbar]
\exp[-N_s \hbar J/{\tilde J}]$ for $J/{\tilde J}> 0.3$, as shown in
Fig.\ref{action}.

This analysis suggests that the metastable configurations are very
stable when many sites must invert their population to reach another
metastable state. However, one should not forget that especially in
larger lattices two metastable configurations might differ just by the
occupation of few lattice sites. This, and the corresponding small
energy difference, should be carefully taken into account in a
realistic analysis at finite temperature.

\subsection{Initialisation and detection}

Very important issues are the initialisation and detection of the
atomic states in the lattice. One can use superlattices in order to
prepare the atoms in configurations of preferential symmetry.
This idea is presently pursued by several experimental groups
\cite{expe}.
We have checked that the presence of defects is strongly reduced when
a local potential energy following desired patterns is added to the
optical lattice. Note that the configurations obtained in such a way
will also remain stable once the superlattice is removed, thanks to
dipole-dipole interaction.

\begin{figure}[h]
\includegraphics[width=12cm,angle=0]{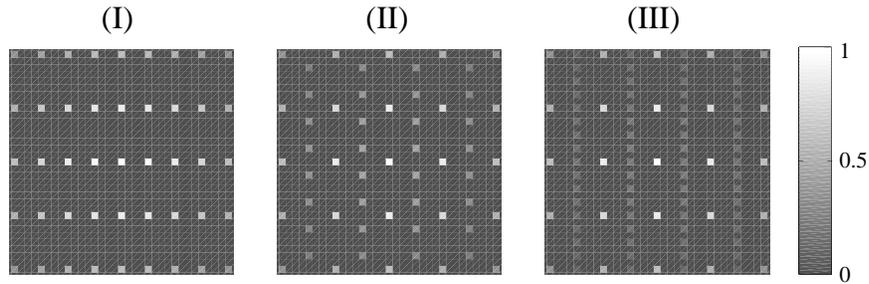}
\caption{Spatial noise correlation patterns for configurations (I) to
(III) in Fig.~\ref{lobes_dipdip}, assuming a localised gaussian
density distribution at each lattice site. Figure from \cite{menotti}.}
\label{fig_fft2}
\end{figure}

The spatially modulated structures created in such a way can be
detected via the measurement of the noise correlations of the
expansion pictures \cite{scarola,altman,bloch3}: the ordered
structures in the lattice give rise to different patterns in the
spatial noise correlation function

\begin{eqnarray}
C(d) = \frac{\int\langle n_{TOF}(x+d/2) n_{TOF}(x-d/2)\rangle \;
d^2 x } {\int \langle n_{TOF}(x+d/2)\rangle \langle
n_{TOF}(x-d/2)\rangle \; d^2x} \approx
\sum_{k,\ell} e^{i(m/\hbar t) d(r_k-r_{\ell})} n_k
n_\ell = |{\cal F} (n)|^2\;  \nonumber
\end{eqnarray}
equal to the modulus square of the Fourier transform of the density
distribution in the lattice ($n_{TOF}$ is the density distribution
after time of flight, while $n_k$ is the density distribution in the
lattice). Such a measurement is in principle able to recognise the
defects in the density distribution, which could be exactly
reconstructed starting from the patterns in the spatial noise
correlation function. For the moment, the signal to noise ratio
required for single defect recognition is beyond the present
experimental possibilities. However, averaging over a finite number of
different experimental runs producing the same spatial distribution of
atoms in the lattice, a good signal can be obtained.
In Fig.\ref{fig_fft2}, we show the noise correlations for the
metastable configurations at filling factor $1/2$ shown in
Fig.\ref{lobes_dipdip}.

Presently we are studying the possibility of transferring in a
controlled way those systems from one configuration to another
\cite{trefzger}.  This, together with the capability of initialising
and reading out the state of the lattice, might make those systems
useful for applications as quantum memories.

\section{Conclusions}

In this paper, we have given a review of some aspects of the physics
of ultra-cold atomic gases interacting via a long-range dipolar
potential. On the experimental side, we have presented an overview of
the state of the art of the experiments, starting from the first
observation of dipolar effects in a Chromium Bose-Einstein condensate
to the most recent experiments demonstrating strong dipolar
interactions in those systems. The description of those experimental
achievements has been accompanied by the explanation of the underlying
theoretical models. On the theoretical side, we have presented our
recent results on dipolar gases in optical lattices, describing the
theoretical framework and explaining in a detailed way the difference
brought in by long-range interactions compared to the case of
zero-range interactions. In particular, we have pointed out that
long-range interactions introduce a huge number of metastable states
in the system, which are not present in the case of zero-range
interaction. We have motivated why Chromium atoms are the best
candidates at the moment for the realisation of such systems.  We hope
that with our work, we have drawn the attention of the statistical
mechanics community devoted to the study of long-range interactions to
this novel kind of systems and that this will be source of inspiration
for future collaborations.

%%%%%%%%%%%%%%%%%%%%%%%%%%%%%%%%%%%%%%%%%%%%%%%%
%% BACKMATTER
%%%%%%%%%%%%%%%%%%%%%%%%%%%%%%%%%%%%%%%%%%%%%%%%

\begin{theacknowledgments}
We thank all the co-authors of our experimental and theoretical papers
on dipolar gases. We acknowledge support by the EU IP Programme
"SCALA", ESF PESC QUDEDIS, MEC (Spanish Government) under contracts
FIS 2005-04627, Consolider Ingeni 2010 ``QOIT'', Acciones Integradas
ICFO-Hannover and by the German Science Foundation (SFB/TRR21 and
Pf381/3-2).  C.M. and T.L. acknowledge financial support by the EU
through an EIF Marie-Curie Action.
\end{theacknowledgments}



%%\bibliographystyle{aipprocl} % if natbib is missing

%\bibliography{sample}

\vspace*{-0.25cm}
\begin{thebibliography}{9}

\bibitem{bec123} H.M. Anderson, J.R. Ensher, M.R. Matthews, C.E. Wieman,
and E.A. Cornell, Science {\bf 269}, 198 (1995);
C.C. Bradley, C.A. Sackett, J.J. Tollett, and R.G. 
Hulet, Phys. Rev. Lett. {\bf 75}, 1687 (1995);
K.B. Davis, M.-O. Mewes, M.R. Andrews, N.J. van Druten,
D.S. Durfee, D.M. Kurn, and W. Ketterle, Phys. Rev. Lett. {\bf 75},
3969 (1995).

\bibitem{jin} B. DeMarco and D.S. Jin, Science {\bf 285},1703 (1999).

\bibitem{coll_osc} D.S. Jin, J.R. Ensher, M.R. Matthews, C.E. Wieman,
and E.A. Cornell, Phys. Rev. Lett. {\bf 77}, 420 (1996); M.-O. Mewes,
M.R. Andrews, N.J. van Druten, D.M. Stamper-Kurn, D.S. Durfee, C.G. 
Townsend, and W. Ketterle, Phys. Rev. Lett. {\bf 77}, 988 (1996).

\bibitem{interference} M.R. Andrews, C.G. Townsend, H.J. Miesner, 
D.S. Durfee, D.M. Kurn, and W. Ketterle, Science {\bf 275}, 637 (1997).

\bibitem{coherence} see e.g.  B.P Anderson and M.A. Kasevich,
Science {\bf 282}, 1686 (1998); I. Bloch, T.W. H\"ansch, and T. Esslinger,
Phys. Rev. Lett. {\bf 82}, 3008 (1999); E.W. Hagley, L. Deng,
M. Kozuma, J. Wen, K. Helmerson, S.L. Rolston, and W.D. Phillips,
Science {\bf 283}, 1706 (1999).

\bibitem{nonlinear} L. Deng, E.W. Hagley, J. Wen,
M. Trippenbach, Y. Band, P.S. Julienne, J.E. Simsarian, K. Helmerson,
S.L. Rolston, and W.D. Phillips, Nature {\bf 398}, 218 (1999).

\bibitem{solitons} S. Burger, K. Bongs, S. Dettmer, W. Ertmer, 
K. Sengstock, A. Sanpera, G. V. Shlyapnikov, and M. Lewenstein, 
Phys. Rev. Lett. {\bf 83}, 5198 (1999);
J. Denschlag, J.E. Simsarian, D.L. Feder, C.W. Clark, 
L.A. Collins,  J. Cubizolles, L. Deng,  E.W. Hagley, 
K. Helmerson,  W.P. Reinhardt, S.L. Rolston,  B.I. Schneider,
W.D. Phillips, Science {\bf 287}, 97 (2000).

\bibitem{sound} M.R. Andrews, D.M. Kurn, H.-J. Miesner, D.S. Durfee, 
C.G. Townsend, S. Inouye, and W. Ketterle, Phys. Rev. Lett. {\bf 79},
553 (1997).

\bibitem{scissor} O.M. Marag\`o, S.A. Hopkins, J. Arlt, E. Hodby, 
G. Hechenblaikner, and C.J. Foot, Phys. Rev. Lett. {\bf 84}, 2056
(2000).

\bibitem{vortices} M.R. Matthews, B.P .Anderson, P.C. Haljan,
D.S. Hall, C.E. Wieman, and E.A. Cornell, Phys. Rev. Lett.
{\bf 83}, 2498 (1999); K.W. Madison, F. Chevy, W. Wohlleben, and
J. Dalibard, Phys. Rev. Lett. {\bf 84}, 806 (2000); J.R. Abo-Shaeer,
C. Raman, J.M. Vogels, and W. Ketterle, Science {\bf 292},
476 (2001).

\bibitem{feshbach} P. Courteille, R.S. Freeland, D.J. Heinzen,
F.A. van Abeelen, and B.J. Verhaar, Phys. Rev. Lett. {\bf 81},
69 (1998); S. Inouye, K.B. Davis, M.R. Andrews, J. Stenger,
H.-J. Miesner, D.M. Stamper-Kurn, and W. Ketterle, Nature
{\bf 392}, 151 (1998).

\bibitem{greiner} M. Greiner, O. Mandel, T. Esslinger, T.W. H\"ansch,
and I. Bloch, Nature {\bf 415}, 39 (2002).

\bibitem{BKT} Z. Hadzibabic, P. Kr\"uger, M. Cheneau, B. Battelier,
and J. Dalibard, Nature {\bf 441}, 1118 (2006).

\bibitem{tonks} T.T. Kinoshita, T. Wenger and D.S. Weiss, Science {\bf 305},
1125 (2004); B. Paredes, A. Widera, V. Murg, O. Mandel, S. F\"olling,
I. Cirac, G.V. Shlyapnikov, T.W. H\"ansch and I. Bloch, Nature {\bf
429}, 277 (2004).

\bibitem{bcsbec} due to the large number of experimental and
theoretical papers on this subject, we address the reader to the
recent review \cite{giorgini}.

\bibitem{molecules} see C. Ospelkaus, S. Ospelkaus, L. Humbert, P. Ernst, 
K. Sengstock, and K. Bongs, Phys. Rev. Lett. {\bf 97}, 120402 (2006) 
and references therein.

\bibitem{baranov}
M.~Baranov, {\L}.~Dobrek, K.~G\'oral, L.~Santos, and
M.~Lewenstein, Physica Scripta {\bf 102}, 74 (2002).

\bibitem{nature} T.~Lahaye, T.~Koch, B.~Fr\"olich, M.~Fattori, J.~Metz,
A.~Griesmaier, S.~Giovanazzi, and T.~Pfau,  Nature {\bf 448},
672 (2007).

\bibitem{menotti} C.~Menotti, C.~Trefzger, and M.~Lewenstein,
Phys. Rev. Lett. {\bf 98}, 235301 (2007).

\bibitem{dalfovo} F.S. Dalfovo, S. Giorgini, L.P. Pitaevskii, and 
S. Stringari, Rev. Mod. Phys. {\bf 71}, 463 (1999).

\bibitem{bec}
L. . Pitaevskii and S. Stringari, {\sl Bose-Einstein condensation},
(Clarendon Press, Oxford, 2003).

\bibitem{pethick} C. Pethick and H. Smith, {\it Bose-Einstein Condensation
in Dilute Gases} (Cambridge University Press, 2002).

\bibitem{bloch} I. Bloch, J. Dalibard, and W. Zwerger,
arXiv:0704.3011.

\bibitem{giorgini} S. Giorgini, L.P. Pitaevskii, and S. Stringari,
arXiv:0706.3360.

\bibitem{EPJD}
Special Issue {\it Ultracold Polar Molecules: Formation and
Collisions}, Ed. J. Doyle, B. Friedrich, R.V. Krems, and
F. Masnou-Seeuws, Eur. Phys. J. D {\bf 31} No.2, (2004).

\bibitem{rydberg} T.F. Gallagher, {\it Rydberg Atoms}
 (Cambridge University Press, 1994); D. Tong, S.M. Farooqi,
 J. Stanojevic, S. Krishnan, Y.P. Zhang, R. C\^{o}t\'{e}, E.E. Eyler
 and P.L. Gould, Phys. Rev. Lett. {\bf 93}, 063001 (2004); T. Vogt,
 M. Viteau, J. Zhao, A. Chotia, D. Comparat and P. Pillet,
 Phys. Rev. Lett. {\bf 97}, 0803003 (2006); R. Heidemann, U. Raitzsch,
 V. Bendkowsky, B. Butscher, R. L\"ow, L. Santos, and T. Pfau,
 Phys. Rev. Lett. {\bf 99}, 163601 (2007).


\bibitem{jurgen}
J. Stuhler, A. Griesmaier, T. Koch, M. Fattori, T. Pfau, 
S. Giovanazzi, P. Pedri, and L. Santos, 
Phys. Rev. Lett. {\bf 95}, 150406 (2005).

\bibitem{axel-phd}
A. Griesmaier, J. Phys. B {\bf 40}, 91 (2007).

\bibitem{clip} P.O. Schmidt, S. Hensler, J. Werner, T. Binhammer, 
A. G\"orlitz and T. Pfau, J. Opt. B: Quantum Semiclass. Opt. {\bf 5},
170 (2003).

\bibitem{doppler} P.O.Schmidt, S.Hensler, J.Werner, 
T.Binhammer, A.G\"orlitz and T.Pfau,
J.Opt.Soc.Am.B {\bf 20}, 5 (2003).

\bibitem{dipolar-relax} S. Hensler, J. Werner, A. Griesmaier, 
P.O. Schmidt, A. G\"orlitz, T. Pfau, S. Giovanazzi and K. Rz\c a{\.z}ewski,
Appl. Phys. B {\bf 77}, 765 (2003).

\bibitem{crbec} A. Griesmaier, J. Werner, S. Hensler, J. Stuhler and 
T. Pfau, Phys. Rev. Lett. {\bf 94}, 160401 (2005).

\bibitem{joerg}
 J. Werner, A. Griesmaier, S. Hensler, J. Stuhler, T. Pfau,
 A. Simoni and E. Tiesinga, Phys. Rev. Lett. {\bf 94}, 183201 (2005).


\bibitem{santosodell} L.~Santos, G.~Shlyapnikov, and M.~Lewenstein,
Phys. Rev. Lett. {\bf 90}, 250403 (2003); see also D.H.~O'Dell,
S.~Giovanazzi, and G.~Kurizki, Phys. Rev. Lett.  {\bf 90}, 110402
(2003).

\bibitem{stefano}
C.Eberlein, S.Giovanazzi, and D.H.J.O'Dell, Phys. Rev. A {\bf 71},
033618 (2004); S. Giovanazzi, P. Pedri, L. Santos, A. Griesmaier,
M. Fattori, T. Koch, J. Stuhler, and T. Pfau, Phys. Rev. A {\bf 74},
013621 (2006).

\bibitem{castindum}
Y. Castin and R. Dum, Phys. Rev. Lett. {\bf 77}, 5315 (1996);
Yu. Kagan, E. L. Surkov, and G. V. Shlyapnikov, Phys. Rev. A {\bf 54}, 
R1753 (1996).

\bibitem{fisher} M.P.A. Fisher, P.B. Weichman, G. Grinstein,
and Daniel S. Fisher, Phys. Rev. B {\bf 40},  546 (1989).

\bibitem{jaksch}  D. Jaksch, C. Bruder, J.I. Cirac, C.W. Gardiner, 
and P. Zoller, Phys. Rev. Lett. {\bf 81}, 3108 (1998).

\bibitem{santos} K. G\'oral, L. Santos and M. Lewenstein,
Phys. Rev. Lett.  {\bf 88}, 170406 (2002).

\bibitem{sinha} D.L.~Kovrizhin,   G.V. Pai, and S. Sinha, Europhys. Lett.
{\bf 72}, 162 (2005).

\bibitem{sengupta} P. Sengupta, L.P.~Pryadko, F.~Alet, M.~Troyer, and
G.~Schmid, Phys. Rev. Lett. {\bf 94}, 207202 (2005).

\bibitem{scarola} V.W.~Scarola, E.~Demler, and S. Das Sarma,
Phys. Rev. A {\bf 73}, 051601(R) (2006).

\bibitem{helium} E.Kim and M.H.W.Chan, Nature {\bf 427}, 225
(2004); Science {\bf 305}, 1941 (2004); Phys.Rev.Lett. {\bf 97},
115302 (2006); S.~Sasaki, F.~Caupin, and S.~Balibar, arXiv:0707.3110.

\bibitem{ashcroft} N.W. Ashcroft and N.D. Mermin, {\it Solid State 
Physics} (Saunders College Publishing, 1976).

\bibitem{sheshadri} K. Sheshadri, H.~R. Krishnamurthy, R. Pandit, and T.V.
Ramakrishnan, Europhys. Lett.{\bf 22},257(1993).

\bibitem{sachdev} Subir Sachdev, {\it Quantum Phase Transitions}
(Cambridge University Press, 1999).

\bibitem{freericks} J.K. Freericks and H. Monien, Europhys. Lett. 
{\bf 26}, 545 (1994); Phys. Rev. B {\bf 53}, 2691 (1996).

\bibitem{qmc} see e.g. R.T. Scalettar, G.G. Batrouni, and G.T. Zimanyi,
Phys. Rev. Lett. {\bf 66}, 3144; W. Krauth, N. Trivedi, and
D. Ceperley, Phys. Rev. Lett. {\bf 67}, 2307 (1991); N.V. Prokofev,
B.V. Svistunov, and I.S. Tupisyn, Phys. Lett. A {\bf 238}, 253 (1998).

\bibitem{batrouni} G. G. Batrouni, V. Rousseau, R. T. Scalettar, 
M. Rigol, A. Muramatsu, P. J. Denteneer, and M. Troyer, Phys. Rev. 
Lett. {\bf 89}, 117203 (2002)

\bibitem{bloch2} S. F\"olling, A. Widera, T. M\"uller, F. Gerbier, 
I. Bloch, Phys. Rev. Lett. 97, 060403 (2006).


\bibitem{tdga} D. Jaksch, V. Venturi, J. I. Cirac, C. J. Williams, 
and P. Zoller, Phys. Rev. Lett. 89, 040402 (2002); J. Zakrzewski,
Phys. Rev. A {\bf 71}, 043601 (2005).


\bibitem{demler} D.-W. Wang, M.D. Lukin, and E. Demler,
Phys. Rev. Lett. {\bf 97}, 180413 (2006).

\bibitem{trefzger} C. Trefzger, C. Menotti and M. Lewenstein,
in preparation.

\bibitem{goral} K.~G\'oral and L.~Santos, Phys. Rev. A {\bf 66},
  023613 (2002).

\bibitem{wen} X.-G.~Wen, {\it Quantum field theory of many-body systems},
(Oxford University Press, Oxford, 2004).

\bibitem{expe} P.J. Lee, M. Anderlini, B.L. Brown, J. Sebby-Strabley, 
W.D. Phillips, and J.V. Porto, Phys. Rev. Lett. {\bf 99}, 020402 (2007);
and I. Bloch, J. Henker-Denschlag, private communications.


\bibitem{altman} E.~Altman, E.~Demler, and M.D.~Lukin, Phys. Rev. A
{\bf 70}, 013603 (2004).

\bibitem{bloch3} S.~F\"olling, F.~Gerbier, A.~Widera, O.~Mandel,
T.~Gericke, and I.~Bloch, Nature {\bf 434}, 481 (2005).

\end{thebibliography}

%  \typeout{******************************************}
%  \typeout{** Please run "bibtex \jobname" to optain}
%  \typeout{** the bibliography and then re-run LaTeX}
%  \typeout{** twice to fix the references!}
%  \typeout{******************************************}

\end{document}